\documentclass[twocolumn, groupedaddress, showpacs, floatfix, letterpaper, 10pt]{revtex4-2}

\usepackage{amsmath, amssymb, amsthm}
\usepackage[english]{babel}
\usepackage[all]{xy}
\usepackage{graphicx}
\usepackage{tikz}
\usetikzlibrary{matrix}
\usepackage{verbatim}
\usepackage{enumerate}
\usepackage{url}
\usepackage{subfigure}
\usepackage{bm}
\usepackage{times}

\usepackage{tikz}
\usetikzlibrary{shapes.geometric,arrows}
\tikzstyle{arrow} = [thick,->,>=stealth]

\newtheorem{prop}{Proposition}
\newtheorem{rem}{Remark}

\newcommand{\note}[1]{}

\newcommand{\SSzero}{{\bf {SS}}_0}
\newcommand{\SSpi}{{\bf {SS}}_\pi}

\usepackage{xcolor}

\definecolor{darkgreen}{rgb}{0.0, 0.5, 0}

\renewcommand{\imath}{i}

\renewcommand{\L}{L}
\newcommand{\Sym}{\mathbf{S}}
\newcommand{\Di}{\mathbf{D}}

\newcommand{\s}{\mathrm{s}}
\newcommand{\n}{\mathrm{n}}

\newcommand{\Sc}{\mathcal{S}}

\newcommand{\Pc}{\mathcal{P}}

\renewcommand{\SSzero}{\boldsymbol{0}}
\renewcommand{\SSpi}{\boldsymbol{\pi}}

\newcommand{\grev}{\gamma^{(t)}}
\newcommand{\Lc}{\mathcal{L}}

\DeclareMathOperator{\Fix}{Fix}

\newcommand{\Sync}{\SSzero}

\newcommand{\ud}{\mathrm{d}}
\newcommand{\udi}{\,\ud}

\newcommand{\R}{\mathbb{R}}

\newcommand{\Z}{\mathbb{Z}}
\newcommand{\N}{\mathbb{N}}

\newcommand{\Tor}{\mathbb{T}}

\newcommand{\X}{\mathbb{X}}

\newcommand{\g}{\gamma}
\newcommand{\G}{\Gamma}

\newcommand{\abs}[1]{\left|#1\right|}

\newcommand{\rset}[2]{\left\lbrace\, #1\,\left|\;#2\right.\right\rbrace}

\newcommand{\set}[2]{\rset{#1}{#2}}

\newcommand{\sset}[1]{\left\lbrace #1\right\rbrace}

\DeclareMathOperator{\divo}{div}
\renewcommand{\div}{\divo}

\begin{document}

\title
{Symmetry breaking yields chimeras in two small populations of Kuramoto-type oscillators} 

\author{Oleksandr Burylko${}^{1}$, Erik A. Martens${}^{2,3}$, and Christian Bick${}^{4,5}$}
\affiliation{
${}^1$Institute of Mathematics, National Academy of Sciences of Ukraine, Tereshchenkivska Str.~3, 01024 Kyiv, Ukraine\\
${}^2$Centre for Mathematical Sciences, Lund University, S\"olvegatan 18, 221 00 Lund, Sweden\\ 
${}^3$Chair for Network Dynamics, Center for Advancing Electronics Dresden (cfaed) and Institute for Theoretical Physics, TU Dresden, 01062 Dresden, Germany\\
${}^4$Department of Mathematics, Vrije Universiteit Amsterdam, De Boelelaan 1111, Amsterdam, the Netherlands\\
${}^5$Department of Mathematics, University of Exeter, Exeter EX4 4QF, United Kingdom
}

\date{\today}

\begin{abstract}
Despite their simplicity, networks of coupled phase oscillators can give rise to intriguing collective dynamical phenomena. 
However, the symmetries of globally and identically coupled identical units do not allow solutions where distinct oscillators are frequency-unlocked---a necessary condition for the emergence of chimeras. 
Thus, forced symmetry breaking is necessary to observe chimera-type solutions. Here, we consider the bifurcations that arise when full permutational symmetry is broken for the network to consist of coupled populations. We consider the smallest possible network composed of four phase oscillators and elucidate the phase space structure, (partial) integrability for some parameter values, and how the bifurcations away from full symmetry lead to frequency-unlocked weak chimera solutions. 
Since such solutions wind around a torus they must arise in a global bifurcation scenario. Moreover, periodic weak chimeras undergo a period doubling cascade leading to chaos. The resulting chaotic dynamics with distinct frequencies do not rely on amplitude variation and arise in the smallest networks that support chaos. 
\end{abstract}

\maketitle 
\allowdisplaybreaks

{\bf
Networks of coupled oscillators occur in a wide range of systems in biology, medicine and technology. 
The proper functioning of such systems often relies on the emergence (or absence) of collective modes such as synchronization, where oscillators lock their frequencies and/or phases~\cite{Strogatz2004,Pikovsky2003}---a phenomenon that can be studied in the prominent Kuramoto model of phase oscillators~\cite{Kuramoto,Acebron2005}.
Chimeras are symmetry breaking solutions where part of the population synchronizes while the other oscillates incoherently~\cite{Kuramoto_Battogtokh_2002}, even if oscillators are identical. While this phenomenon has received much attention in recent years~\cite{Panaggio2016,Omelchenko2018}, the bifurcations of chimera solutions in finite networks have remained elusive. 
In physical/mechanical oscillator models, the emergence of chimera states can be related to resonance~\cite{MartensThutupalli2013,MartensThutupalli2022}, and in phase oscillator models to phase lags between oscillators being near $\pm\frac{\pi}{2}$~\cite{Ashwin2014a,Martens2016}. Here, we analyze a variant of the Kuramoto model with two populations with two phase oscillators near these parameter points. 
We present a detailed how phase space is organized and elucidate the emergence and bifurcations of chaotic weak chimeras~\cite{Bick2017,Martens2016,Bick2018}.
}

\section{Introduction}

\noindent
Coupled phase oscillator networks have been instrumental in understanding collective dynamic phenomena in real-world oscillatory systems~\cite{Pikovsky2003,Strogatz2004}. To understand the dynamics, in many cases it makes sense to assume an idealized system where all nodes are identically connected to all other nodes (such as in the Kuramoto model~\cite{Kuramoto,Acebron2005}) and/or all nodes are identical (for example, in the classical master stability approach~\cite{Pecora1998}). If both of these assumptions are made for a network of $\L\in\N$ oscillators, then the resulting dynamical system is $\Sym_\L$-equivariant, where~$\Sym_\L$ denotes the group of permutations of~$\L$ symbols. While these systems can still exhibit intriguing dynamics~\cite{Bick2011, Pikovsky2015a}, the symmetry particularly restricts the speed at which the oscillators evolve: For networks of phase oscillators, all nodes have to be frequency-synchronized since oscillators' phases cannot pass each other~\cite{Ashwin2014a}.

Dynamics where different oscillators evolve at different (asymptotic) frequencies have attracted significant attention in recent years, a phenomenon commonly known as a \emph{chimera}; see Refs.~\onlinecite{Panaggio2015,Schoell2016,Omelchenko2018} for reviews.
As such solutions/invariant sets are impossible in $\Sym_\L$-symmetric phase oscillator networks, breaking the full permutational symmetry is necessary for frequency unlocked solutions to arise. To break the symmetries, one can for example fix a proper subgroup $H\subset \Sym_\L$ and then consider perturbations to the dynamical equations such that the perturbed system is $H$-equivariant---this is also known as \emph{forced symmetry breaking}~\cite{Guyard1999}. Breaking the symmetries will break some of the invariant subspaces that restrict the dynamics and create the possibility for frequency-unlocked chimera-type solutions. For example, consider perturbing a network of $\L=MN$ oscillators to form~$M$ populations of~$N$ oscillators each: Let $\theta_{\sigma, k}\in\Tor:=\R/2\pi\Z$ 
denote the phase of  oscillator~$k\in\sset{1, \dotsc, N}$ of population $\sigma\in\sset{1, \dotsc, M}$. The phase  evolves according to
\begin{equation}\label{eq:PopMN}
\begin{split}
\dot\theta_{\sigma,k}  := \frac{\ud}{\ud t}\theta_{\sigma,k} &= \omega + \frac{1}{MN}\sum_{j=1}^N g_\s(\theta_{\sigma,j}-\theta_{\sigma,k})\\ &\qquad\quad+ \frac{1}{MN}\sum_{\tau\neq\sigma}\sum_{j=1}^N g_\n(\theta_{\tau,j}-\theta_{\sigma,k}),
\end{split}
\end{equation}
where $g_\s:\Tor\to\R$ determines the \emph{self-coupling} within populations and $g_\n:\Tor\to\R$ determines the \emph{coupling to other populations} (i.e., `neighbors').
The system has a wreath product symmetry~$H$, whose elements act within populations or permute populations~\cite{Dionne1996a}.
Note that, by definition, the perturbed system retains the symmetries~$H$; thus the type of symmetry breaking is different from adding heterogeneity~\cite{Ashwin2006}, which, generically, does not preserve any symmetries.
In particular, the populations in~\eqref{eq:PopMN} are interchangeable in contrast to networks with distinct populations~\cite{Hong2011a}.

Splitting the oscillators into just $M=2$ populations can lead to frequency-unlocked chimera dynamics. Indeed, the classical work by Abrams {\it et al.}~(Ref.~\onlinecite{Abrams2008}) demonstrated the emergence of chimera solutions for two coupled populations with disparate coupling strength (but identical coupling functions $g_\s=g_\n$) in the limit of $N\to\infty$ oscillators. More generic interactions with nonidentical phase lags and a single harmonic in the limit of large networks may lead to chaotic collective dynamics~\cite{Martens2016,Bick2018}. Chimera solutions not only arise in the mean-field limit but also for finite networks~\cite{Panaggio2016} and general sinusoidal coupling is sufficient to obtain chaotic dynamics~\cite{Bick2018}. While it was already indicated in Ref.~\onlinecite{Bick2018} that chaotic dynamics are possible even in the smallest networks of $M=2$ populations of $N=2$ oscillators, the bifurcations that lead to chaotic chimera dynamics remained elusive.

\begin{figure}
  \raisebox{6.1cm}{\mbox{(a)}}\hspace*{-0.3cm}
\centerline{
\newcommand\XA{-3.7}
\newcommand\YA{-2.6}
\newcommand\YFig{2.65}
\begin{tikzpicture}
  \draw (0, 0) node[inner sep=0] {\includegraphics[width=\columnwidth]{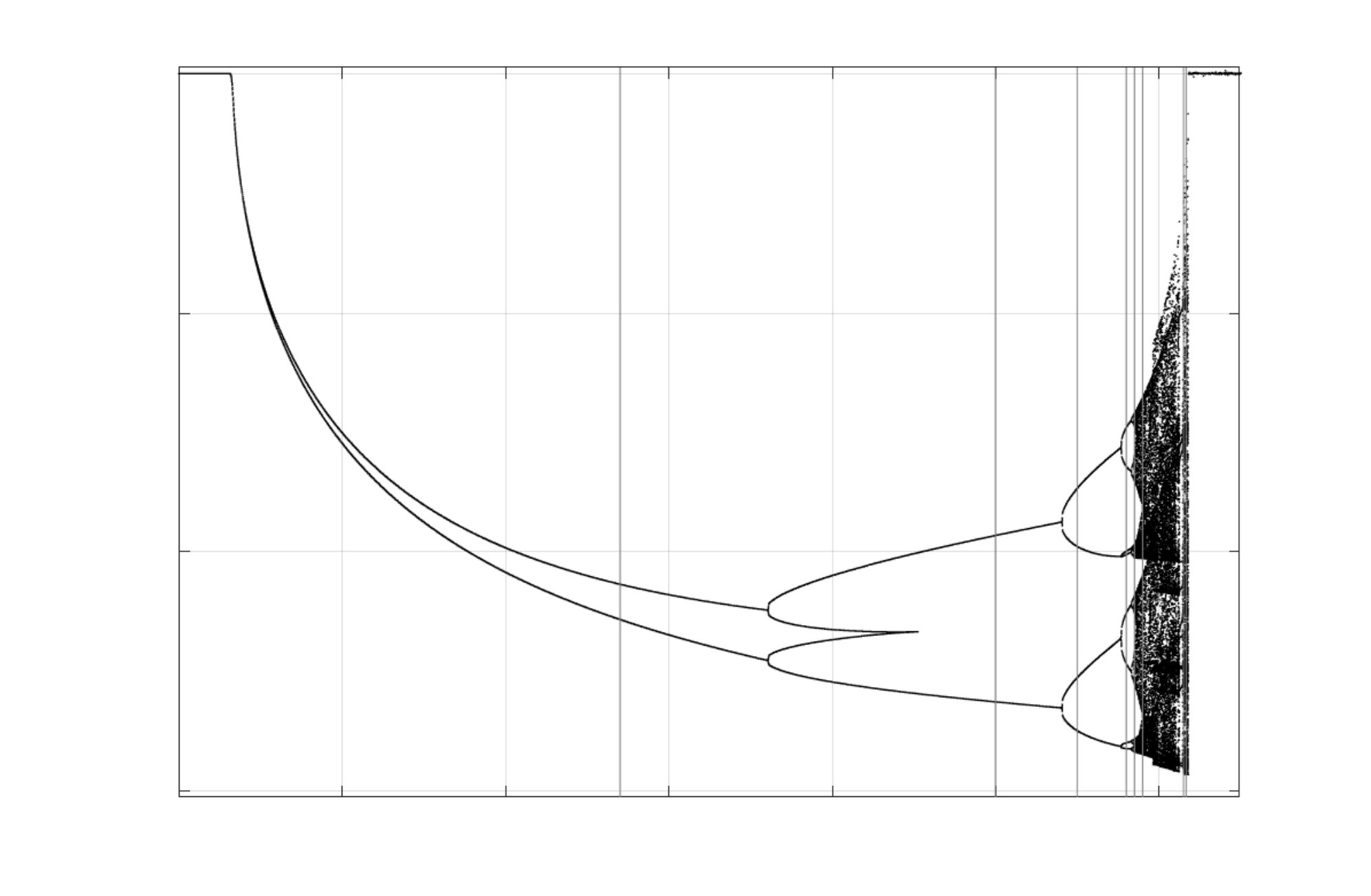}};
  \node[rotate=0] at (0, \YA-0.65) {$\alpha_\s$};
  \node[rotate=90] at (\XA-0.8, 0) {$\min_t{\psi_1(t)},\max_t{\psi_1(t)}$}; 
  \node[rotate=0] at (-3.1, \YA) {1.58};
  \node[rotate=0] at (-1.05, \YA) {1.60};
  \node[rotate=0] at (1.05, \YA) {1.62};
  \node[rotate=0] at (3.1, \YA) {1.64};  
  \newcommand\zy{2.4}
  \newcommand\zh{0.2}
  \newcommand\zhh{0.3}
  \newcommand\asone{-0.41}
  \newcommand\astwo{1.96}
  \newcommand\asthree{2.48}
  \newcommand\asfour{2.79}
  \newcommand\asfive{2.84}
  \newcommand\assix{2.885}
  \newcommand\asseven{3.145}
    \newcommand\aseight{3.165}

  \node[rotate=0] at (\XA + 0.2, 2.4) {$0$};
  \node[rotate=0] at (\XA, 0.8) {$-\dfrac{\pi}{4}$};
  \node[rotate=0] at (\XA, -0.65) {$-\dfrac{\pi}{2}$};
  \node[rotate=0] at (\XA, -2.18) {$-\dfrac{3\pi}{4}$};
  \node[rotate=0] at (-0.6, \YFig) {\footnotesize (a)};
  \node[rotate=0] at (1.8, \YFig) {\footnotesize (b)};
  \node[rotate=0] at (2.3, \YFig) {\footnotesize (c)};
  \node[rotate=0] at (\asfour-0.3, \YFig+\zh) {\footnotesize (d)};
  \node[rotate=0] at (\asfive-0., \YFig+\zhh+0.05) {\footnotesize  (e)};
  \node[rotate=0] at (\assix+0.2, \YFig+\zh) {\footnotesize (f)};
  \node[rotate=0] at (\asseven+0.2, \YFig+\zhh) {\footnotesize (g)};
  \node[rotate=0] at (\aseight+0.4, \YFig+0.1) {\footnotesize (h)};
  \draw (\asone,\zy) -- (\asone,\zy+\zh);
  \draw (\astwo,\zy) -- (\astwo,\zy+\zh);
  \draw (\asthree,\zy) -- (\asthree,\zy+\zh);
  \draw (\asfour,\zy) -- (\asfour,\zy+\zh) -- (\asfour-0.2,\zy+\zhh);
  \draw (\asfive,\zy) -- (\asfive,\zy+\zhh) -- (\asfive,\zy+\zhh+0.15);
  \draw (\assix,\zy) -- (\assix,\zy+\zh) -- (\assix+0.2,\zy+\zhh);
  \draw (\asseven,\zy) -- (\asseven,\zy+\zh) -- (\asseven+0.1,\zy+\zhh);
  \draw (\aseight,\zy) -- (\aseight,\zy+\zh) -- (\aseight+0.2,\zy+\zhh);
  \end{tikzpicture}}\\
  \raisebox{5.1cm}{\mbox{(b)}}\hspace*{-0.3cm}\includegraphics[width=\columnwidth]{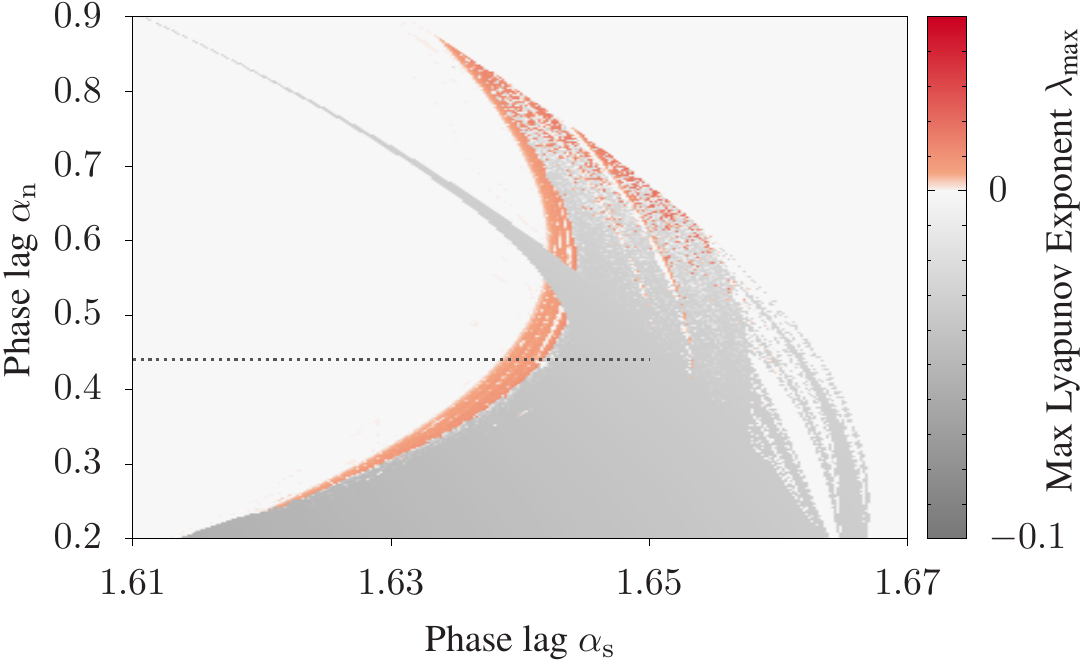}
\caption{\label{fig:BifDiagChaos}The road to chaotic weak chimeras. 
Panel~(a) shows a bifurcation diagram obtained via quasi-continuation (see Sec.~\ref{sec:chaoticweakchimeras}) for fixed $A=0.7$ and $\alpha_\s=0.44$. Vertical lines delineate $\alpha_\s$-values for the trajectories shown later in Fig.~\ref{fig:B-M_Chaos-Proj}, referring to the corresponding panel labels.
Panel~(b) shows the maximal Lyapunov exponent for varying $(\alpha_\s, \alpha_n)$ calculated by numerically integrating from a fixed initial condition for $T=10000$ time units. The dashed line indicates parameter values shown in~(a).}
\end{figure}
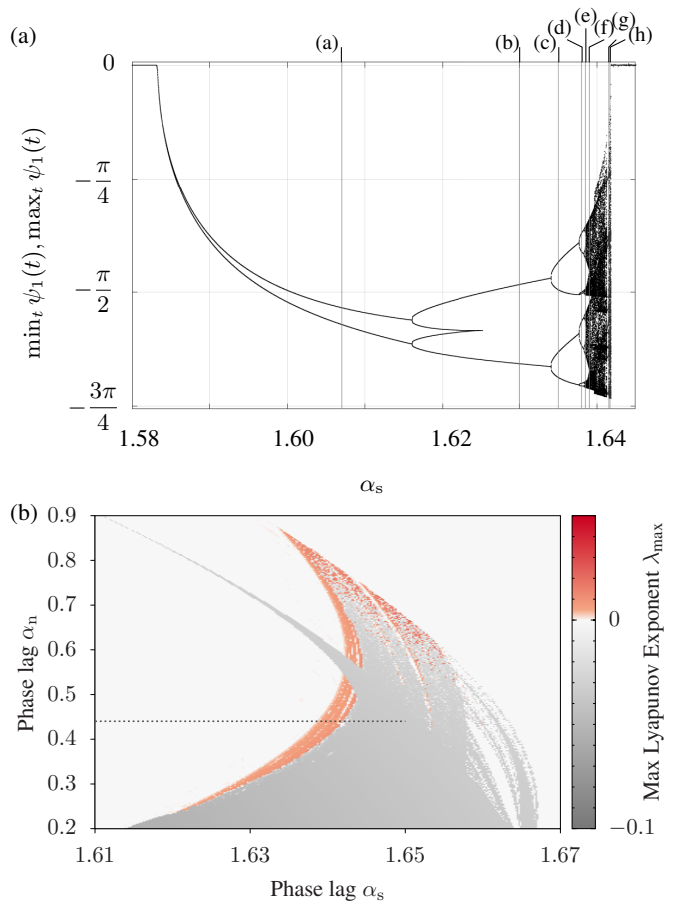

In this paper, we analyze symmetry breaking away from full permutational symmetry to understand the dynamics of small networks of $M=2$ populations of $N=2$ oscillators with generic sinusoidal coupling. This means that while the coupling functions~$g_\s, g_\n$ both have a single harmonic, they may have with distinct phase shifts and amplitudes. We first analyze degenerate cases where the dynamical system is either a gradient system or Hamiltonian-like with conserved quantities. These {special} cases allow to understand how frequency unlocked dynamics arise in these small networks: Perturbing away from the singular parameter values yields global bifurcation scenarios that give rise to frequency-unlocked limit cycle solutions---they correspond to \emph{weak chimeras}.
As such, these solutions have nonzero winding number, wrap around the torus, and have nontrivial homology.
How (some of) these solutions bifurcate \footnote{Note that further branches co-exist.} further is shown in the numerical bifurcation diagram in Fig.~\ref{fig:BifDiagChaos}(a): The dynamics eventually undergo a period doubling cascade that leads to chaotic weak chimeras. Finally, we highlight that for certain parameter values, the system has simultaneously conservative and dissipative dynamics in different regions of phase space.

The paper is organized as follows. In Sect.~\ref{sec:Prelims} we collect basic information about the governing equations~\eqref{eq:PopMN}, identify invariant subspaces and equilibria, and analyze their linear stability. In Sect.~\ref{sec:GradientDyn}, we consider the case of pure sine coupling (\emph{odd coupling}) for which the equations of motion are a gradient system. In Sect.~\ref{sec:ConservativeDyn}, we consider the case of pure cosine coupling (\emph{even coupling}) that leads to the emergence of conserved quantities; the phase space structure elucidates the emergence or frequency-unlocked solutions 
{corresponding to weak chimeras}. Subsequently, in Sect.~\ref{sec:DissipativeDyn} we show how these weak chimeras bifurcate. Specifically, we detail the bifurcations in Fig.~\ref{fig:BifDiagChaos} that lead to the emergence of chaotic dynamics. In Sect.~\ref{sec:ConservativeDissipative}, we briefly comment on coexisting conservative dissipative dynamics for combined pure sine/cosine coupling. We conclude with a discussion in the final section.

\section{From global coupling to coupled populations of Kuramoto oscillators}
\label{sec:Prelims}

Before we analyze coupled phase oscillator networks, we briefly recall some notions related to dynamical systems that are equivariant with respect to the action of a group of symmetries. Let $F:\X\to T\X$ be a smooth vector field on~$\X$ where~$T\X$ denotes the tangent
bundle. Suppose that a group~$\Gamma$ acts on~$\X$. The vector field~$F$
is $\G$-equivariant if
\begin{equation*}
F(\gamma x) = \hat{\gamma}F(x)
\end{equation*}
for all $\gamma\in\Gamma$ where~$\hat{\gamma}$ is the induced
action on the tangent space. A $\G$-equivariant vector field~$F$ defines
a \emph{$\Gamma$-equivariant dynamical system}
\begin{equation}
\label{eq:EqDynSyst}
\dot x = F(x)
\end{equation}
on~$\X$~\cite{Golubitsky2002, Field2007a}. The group $\G_x := \set{\g\in\G}{\g x = x}$
denotes the \emph{stabilizer} or \emph{isotropy subgroup} of $x\in\X$.
While the isotropy subgroup describes the symmetries of a single point, the \emph{symmetries of a set $X\subset\X$} are $\Sigma(X):=\set{\g\in\G}{\g X \subset X}$ where $\g X = \set{\g x}{x\in X}$.
If $H\subset\Gamma$ is a subgroup, then the set $\Fix(H) := \set{x\in\X}{\gamma x = x\ \forall \gamma\in H}$ is invariant under the flow induced by~\eqref{eq:EqDynSyst}.

\subsection{Networks of~$\L$ oscillators}
First consider a network dynamical system~\eqref{eq:PopMN} with $\L\in\N$ phase oscillators. For the remainder of the paper, we will assume that the coupling is of Kuramoto-type as in Ref.~\onlinecite{Martens2016,Choe2016,Bick2018}, i.e., the coupling functions only contain the first harmonic. Specifically, we have
$g_\s(\phi)=K_\s\sin(\phi-\alpha_\s)$,
$g_\n(\phi)=K_\n\sin(\phi-\alpha_\n)$,
where~$K_\s, K_\n>0$ are the coupling strengths within and between populations and $\alpha_\s, \alpha_\n\in\Tor$ the phase lags.
By rescaling time we can set $K_\s+K_\n=1$ without loss of generality and parametrize the coupling strength through the disparity parameter $A:=K_\s-K_\n$; see also Ref.~\onlinecite{Abrams2008}. Note that since the coupling is through phase differences, the system is~$\Tor$ equivariant: There is a continuous~$\Tor$ symmetry where $\gamma\in\Tor$ acts on the phase space $\Tor^\L$ by a common phase shift $\theta_{\sigma,k}\mapsto\theta_{\sigma,k}+\gamma$ to all oscillators.

\newcommand{\CIR}{\mathcal{C}}

\subsubsection{Globally and identically coupled oscillators}
\label{sec:MNGlobal}

If the coupling within and between populations is identical, that is, $A=0$ ($K_\s=K_\n$) and $\alpha_\s=\alpha_\n=:\alpha$, then we have a globally coupled network of $\L=MN$ identical oscillators. Omitting the population index~$\sigma$, write $\theta=(\theta_1, \dotsc, \theta_\L)$ and the evolution of the phase~$\theta_k$ of oscillator $k\in\sset{1, \dotsc, \L}$ is
\begin{equation}\label{eq:KuramotoSym}
\dot\theta_k = \omega + \frac{1}{\L}\sum_{j=1}^\L\sin(\theta_j-\theta_k-\alpha).
\end{equation}
The system~\eqref{eq:KuramotoSym} is~$\Sym_\L$-equivariant, where~$\Sym_\L$ denotes the symmetric group on~$\L$ symbols which acts on~$\Tor^\L$ by permuting the oscillator indices. This implies that the sets $\Pc_{mn} := \set{\theta}{\theta_m=\theta_n}$ are dynamically invariant as fixed-point sets of the transposition that swaps oscillators~$m$ and~$n$. Thus, the dynamics can be reduced to the canonical invariant region
\[\CIR = \sset{\theta_1<\theta_2<\dotsb<\theta_\L<\theta_1+2\pi}\]
that is bounded by~$\Pc_{nm}$; see Refs.~\onlinecite{Ashwin1992,Ashwin2016} for details. 
Indeed, since the coupling function only contains a first harmonic, the canonical invariant region is foliated into dynamically invariant sheets on which the dynamics are effectively 2-dimensional~\cite{Watanabe1994}.

\newcommand{\Ip}{\mathrm{I}}
\newcommand{\Ic}{\mathcal{I}}

The symmetries and type of coupling imply that certain phase configurations are dynamically invariant. The phase configuration
\begin{equation}
\Sync = \set{\theta}{\theta_1=\theta_2 = \dotsb = \theta_\L},
\end{equation}
where all oscillators are \emph{phase-synchronized} is dynamically invariant as the intersection of all $\Pc_{mn}$. Phase synchrony is typically quantified by the \emph{Kuramoto order parameter}
\begin{equation}\label{eq:KuramOP}
Z(\theta) = \frac{1}{\L}\sum_{j=1}^\L\exp(i\theta_j)
\end{equation}
with $i:=\sqrt{-1}$:
We have $\abs{Z} = 1$ if and only if $\theta\in\Sync$.
Define the \emph{antiphase} or \emph{incoherent phase configurations} as
\begin{equation}
\Ic = \set{\theta}{Z(\theta)=0}.
\end{equation}
The set~$\Ic$ is dynamically invariant for globally coupled networks and is a union of manifolds (see Ref.~\onlinecite{Ashwin2016}).

The symmetries also constrain the frequencies of the oscillators. For a solution~$\theta(t)$ of~\eqref{eq:KuramotoSym} with initial condition $\theta(0)=\theta^0$ define the \emph{asymptotic average frequency}\cite{Bick2015c,Bick2015d} as
\begin{equation}
\Omega_k(\theta^0) = \lim_{T\to\infty}\frac{1}{T}\int_0^T\dot\theta_k(t)\udi t.
\end{equation}
Since the sets~$\Pc_{mn}$ are dynamically invariant, the phase difference between oscillators is bounded. This implies that
\begin{equation}\label{eq:FreqFullSym}
\Omega_m(\theta^0) = \Omega_n(\theta^0)
\end{equation}
for all $m,n\in\sset{1, \dotsc, \L}$. Recall that a distinguishing feature of a \emph{weak chimera} is the separation of frequencies~\cite{Ashwin2016}:
A trajectory~$\theta(t)$ with initial condition $\theta(0)=\theta^0$ converges to a weak chimera if there are distinct $m,n,n'$ such that $\Omega_m(\theta^0)\neq\Omega_n(\theta^0)=\Omega_{n'}(\theta^0)$. Thus, weak chimeras cannot exist in fully symmetric systems.

\subsubsection{Breaking full symmetry: $M$~populations of $N$~oscillators}
\label{sec:MNPop}

Breaking the symmetry is necessary to observe weak chimeras in a network of phase oscillators: Here, we consider the dynamics~\eqref{eq:PopMN} of~$M$ identical populations of~$N$ oscillators each. Recall that $\L=MN$. Write $\theta_\sigma=(\theta_{\sigma,1}, \dotsc, \theta_{\sigma,N})\in\Tor^N$ for the state of population $\sigma\in\sset{1,\dotsc,M}$ and $\theta=(\theta_1, \dotsc, \theta_M)\in\Tor^{\L}$ for the state of the entire network.
For the Kuramoto-type coupling we are considering, the phase~$\theta_{\sigma,k}$ of oscillator~$k$ in population~$\sigma$ evolves according to
\begin{equation}\label{eq:KuramotoMN}
\begin{split}
\dot\theta_{\sigma,k} &= \omega + \frac{K_\s}{MN}\sum_{j=1}^N \sin(\theta_{\sigma,j}-\theta_{\sigma,k}-\alpha_\s) 
\\&\qquad\quad+ \frac{K_\n}{MN}\sum_{\tau\neq\sigma}\sum_{j=1}^N \sin(\theta_{\tau,j}-\theta_{\sigma,k}-\alpha_\n).
\end{split}
\end{equation}
These equations are $\Sym_N \wr \Sym_M = (\Sym_N)^M\rtimes \Sym_M \subset \Sym_\L$ equivariant where $\rtimes$ denotes a semidirect product. Each copy of~$\Sym_N$ acts on~$\Tor^\L$ by permuting the oscillator indices within a given population~$\sigma$ and~$\Sym_M$ acts by permuting the populations (i.e., the population indices~$\sigma$); the symmetries within and between populations do not necessarily commute and hence the semi-direct product.

\newcommand{\Xp}{\mathrm{X}}

The system has fewer symmetries than the fully symmetric system. For example---in the notation of the previous section---the sets~$\Pc_{mn}$ are only dynamically invariant if oscillators $m,n$ belong to the same population. Note that the oscillators in each population are still fully symmetric. This implies that for \eqref{eq:KuramotoMN} there are invariant sets $\Pc_{\sigma,mn} := \set{\theta}{\theta_{\sigma,m}=\theta_{\sigma,n}}$ for $m,n\in\sset{1,\dotsc,N}$ and any $\sigma\in\sset{1,\dotsc,M}$. 

The (global) Kuramoto order parameter~$Z$ is defined as in~\eqref{eq:KuramOP}; it quantifies coherence of all oscillators in the network. Naturally, we also define phase synchrony and incoherence for each population: The (local) Kuramoto order-parameter for population~$\sigma$ is
\begin{equation}\label{eq:KuramOPpop}
Z_\sigma(\theta) = \frac{1}{N}\sum_{j=1}^N\exp(i\theta_{\sigma,j})
\end{equation} 
and we have $Z = \frac{1}{M}\sum_{\sigma=1}^M Z_\sigma$.

\subsection{Networks of~$4$ oscillators}
\label{sec:2x2}

\begin{figure}
\begin{center}
 \includegraphics[width=\linewidth]{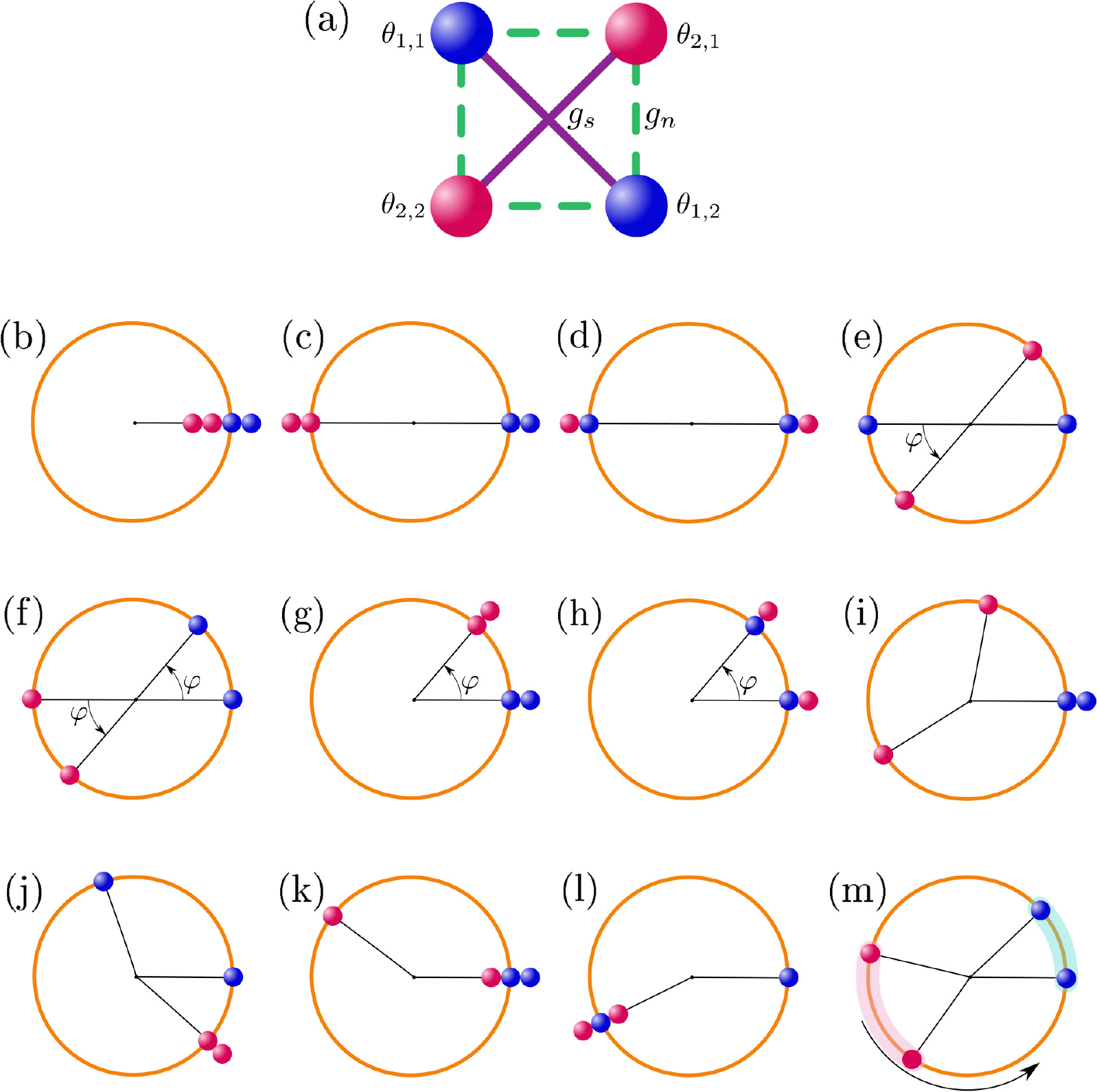}
 \end{center}
 \caption{\label{fig:B-M_Ph-Config}
 A phase oscillator network of $M=2$ populations of $N=2$ oscillators is shown in Panel~(a); the system has the symmetries of the square~$\Di_4$.
 Panels (b--m) show different phase configurations: 
 (b)~full synchrony~$\SSzero$
 (c)~antiphase clusters~$\SSpi$, 
 (d)~antiphase configurations~$(\pi, 0, \pi)$ and~$(\pi, \pi, \pi)$,
 (e)~set~$\Ic_0$ (includes~(d)),
 (f)~set~$\Ic_1$ and~$\Ic_2$,
 (g)~set~$\Sc_0$,
 (h)~sets~$\Sc_1$ and~$\Sc_1$,
 (i)~cluster with $r_1=1$ ($\psi_1=0$) (includes flat chimera),
 (j)~cluster with $r_2=1$ ($\psi_3=0$) (includes flat chimera),
 (k,l)~two-cluster configuration (isotropy $\Sym_3\times\Sym_1$ in the globally coupled system),
 (m) schematic representation of the weak chimera states when oscillators inside each cluster are phase locked ($\abs{\theta_{k,1}-\theta_{k,2}}<2\pi$ for all times) and two groups are phase unlocked (rotate relative to one another).
}
\end{figure}

For the remainder of the section, we consider particular small oscillator networks that consist of $M=2$ populations of $N=2$ oscillators; the network is sketched in Fig.~\ref{fig:B-M_Ph-Config}(a).
In this case, the governing equations~\eqref{eq:KuramotoMN} read
\begin{subequations}\label{eq:goveqfull}
\begin{align}
\begin{split}
\dot\theta_{1,1} &= \omega'+\frac{1}{4}
\big(
 K_\s \sin{(\theta_{1,2}-\theta_{1,1} -\alpha_\s )}\\&\qquad\qquad\qquad
+K_\n \sin{(\theta_{2,1}-\theta_{1,1} -\alpha_\n )}\\&\qquad\qquad\qquad
+K_\n \sin{(\theta_{2,2}-\theta_{1,1} -\alpha_\n )}
\big)
\end{split}\\
\begin{split}
\dot\theta_{1,2} &= \omega'+\frac{1}{4}
\big(
 K_\s \sin{(\theta_{1,1}-\theta_{1,2} -\alpha_\s )}\\&\qquad\qquad\qquad
+K_\n \sin{(\theta_{2,1}-\theta_{1,2} -\alpha_\n )}\\&\qquad\qquad\qquad
+K_\n \sin{(\theta_{2,2}-\theta_{1,2} -\alpha_\n )}
\big)
\end{split}\\
\begin{split}
\dot\theta_{2,1} &= \omega'+\frac{1}{4}
\big(
 K_\s \sin{(\theta_{2,2}-\theta_{2,1} -\alpha_\s )}\\&\qquad\qquad\qquad
+K_\n \sin{(\theta_{1,1}-\theta_{2,1} -\alpha_\n )}\\&\qquad\qquad\qquad
+K_\n \sin{(\theta_{1,2}-\theta_{2,1} -\alpha_\n )}
\big)
\end{split}\\
\begin{split}
\dot\theta_{2,2} &= \omega'+\frac{1}{4}
\big(
 K_\s \sin{(\theta_{2,1}-\theta_{2,2} -\alpha_\s )}\\&\qquad\qquad\qquad
+K_\n \sin{(\theta_{1,1}-\theta_{2,2} -\alpha_\n )}\\&\qquad\qquad\qquad
+K_\n \sin{(\theta_{1,2}-\theta_{2,2} -\alpha_\n )}
\big)
\end{split}
\end{align}
\end{subequations}
with $\omega'=\omega-\frac{K_\s}{4}\sin\alpha_\s$.
If the coupling is fully symmetric, i.e., $K_\s=K_\n$ and $\alpha_\s=\alpha_\n$, then~\eqref{eq:goveqfull} is $\Sym_4$-equivariant. That means that up to the rotational symmetry, the canonical invariant region is a tetrahedron bounded by the dynamically invariant~$\Pc_{mn}$ (as in Sect.~\ref{sec:MNGlobal}); see~Refs.~\onlinecite{Burylko-Pikovsky,Ashwin2016} for details.
The phase configuration~$\SSzero$ with full phase synchrony corresponds to each of the four corners of the tetrahedron. The incoherent phase configurations~$\Ic$ in~$\CIR$ are the points with $\Sym_2\times \Sym_2$ isotropy, that is, a line of points parametrized by $(a, b, a+\pi, b+\pi)\in \Tor^4$.

For a generic choice of coupling parameters, the dynamical system~\eqref{eq:goveqfull} is $\Sym_2\wr \Sym_2\equiv \Di_4$-equivariant, the symmetries of the square; cf.~Fig.~\ref{fig:B-M_Ph-Config}(a). Specifically, $\Di_4$ is generated by the rotational symmetry
\begin{subequations}\label{eq:SymGen}
\begin{align}
\overline{\gamma}_\mathrm{r} &: (\theta_{1,1},\theta_{1,2},\theta_{2,1},\theta_{2,2}) \mapsto (\theta_{2,2},\theta_{2,1},\theta_{1,1},\theta_{1,2}),
\intertext{(a clockwise rotation by an angele of~$\frac{\pi}{2}$ in Fig.~\ref{fig:B-M_Ph-Config}(a)) and the mirror symmetry}
\overline{\gamma}_\mathrm{m}&: (\theta_{1,1},\theta_{1,2},\theta_{2,1},\theta_{2,2}) \mapsto (\theta_{1,1},\theta_{1,2},\theta_{2,2},\theta_{2,1})
\end{align}
\end{subequations}
(a flip about the main diagonal in Fig.~\ref{fig:B-M_Ph-Config}(a)).
General properties of $\Di_4$-equivariant phase oscillator networks can be found in Ref.~\onlinecite{Ashwin1992}.
The codimension-1 invariant subspaces are $\Pc_{1,12}=\sset{\theta_{1,1}=\theta_{1,2}}$ and $\Pc_{2,12}=\sset{\theta_{2,1}=\theta_{2,2}}$. Since we only have two oscillators per population we will write $\Pc_{1} := \Pc_{1,12}$, $\Pc_{2} := \Pc_{2,12}$ for simplicity.

\begin{figure}
\begin{center}
 \includegraphics[width=\linewidth]{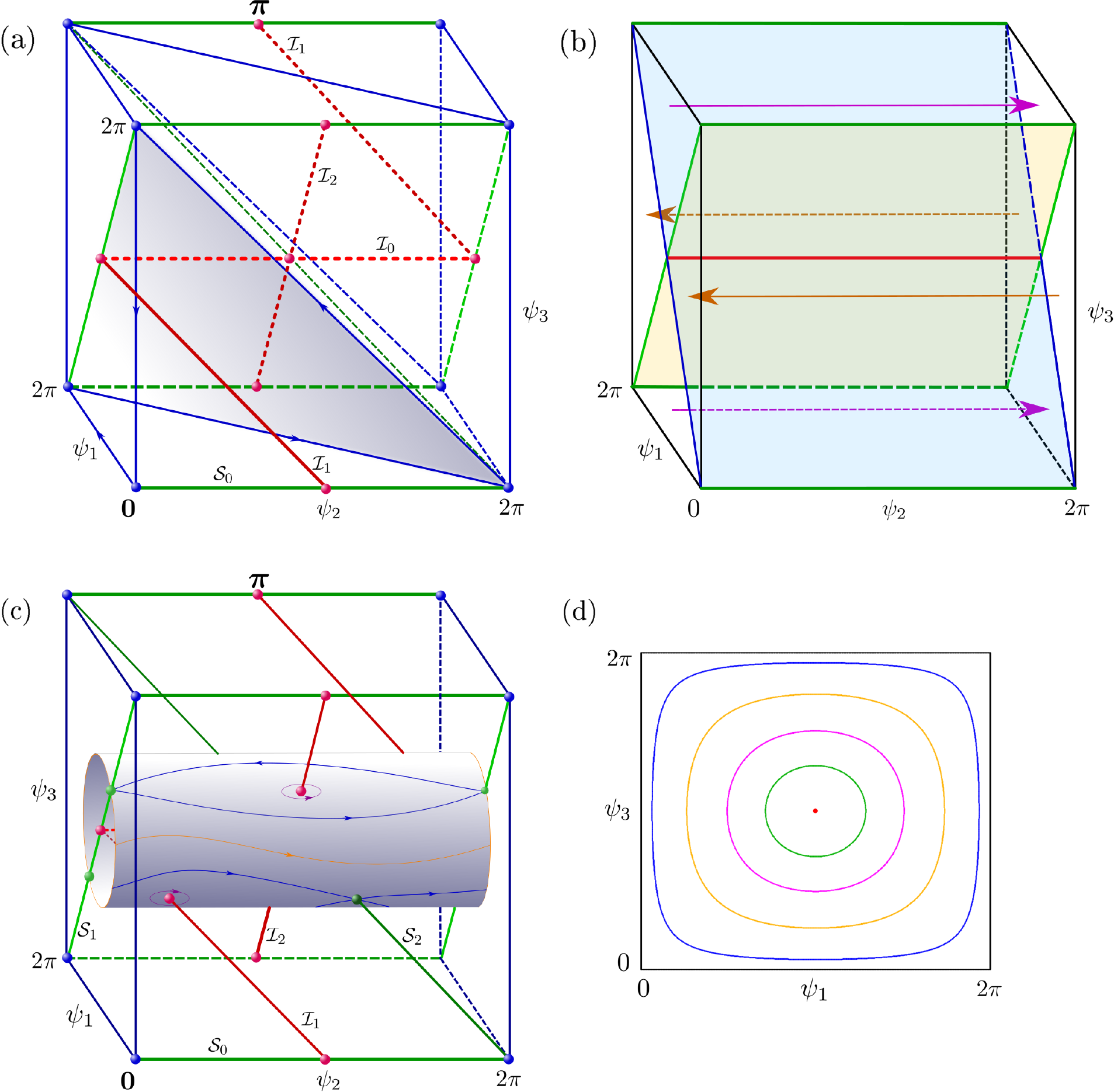}
 \end{center}
 \caption{\label{fig:Cube_Psi-var}
The phase space of the coupled oscillator network~\eqref{eq:goveqn} for $(\psi_1,\psi_2,\psi_3)\in\Tor^3$ is organized by invariant manifolds.
Panel~(a) shows the invariant lines for identical Kuramoto--Sakaguchi coupling $K_\s=K_\n=K$ ($A=0$) and $\alpha_\s=\alpha_\n=\alpha$.
The sets~$\Ic_0$,~$\Ic_1$,~$\Ic_2$ are shown as red lines,
the sets~$\Sc_0$, $\Sc_1$, $\Sc_2$ with isotropy $\Sym_2\times\Sym_2$ as green lines,
and blue lines are points of isotropy $\Sym_3\times\Sym_1$.
Blue balls indicate synchronous state $\SSzero=(0,0,0)$ and its copies, red balls shows the fixed point $\SSpi=(0,\pi,0)$ as well as the points $(\pi,0,\pi)$, $(\pi,\pi,\pi)$ with isotropy $(\Sym_2)^2\times\Z_2$ (and their copies). 
The last fixed points locate on intersections of~$\Ic_\ell$. Light grey plane together with coordinate planes $\psi_\ell=0$ forms one of canonical invariant regions (CIR).
Panel~(b) shows the phase space~$\Tor^3$ for the uncoupled groups $K_\n=0$ ($A=1$) and $\alpha_\s=\pm\frac{\pi}{2}$. The planes $\psi_3=\psi_1$ and $\psi_3=-\psi_1$ are filled with fixed points. These two planes split phase space into two regions that are filled with straight lines (phase-unlocked periodic orbits) oriented in two opposite directions.
Panel~(c) gives a schematic diagram of the relative position of invariant lines and one of the invariant cylinders~$\Lc(C)$ (cf.~\eqref{eq:Cylinders}) for the Hamiltonian-like case $\alpha_\s=\alpha_\n=\pm\frac{\pi}{2}$, $A\in (0,1)$. Blue lines on the cylinder correspond to a heteroclinic (homoclinic on the torus) network. Orange line shows one of continuous family phase uncoupled periodic orbits  that correspond chimeras with nontrivial winding number. There are two chimera strips of periodic orbits bounded by heteroclinic chimeras that have opposite orientation along variable~$\psi_2$. 
Panel~$(d)$ shows the projection of cylinders~$\Lc(C)$ into the plane~$(\psi_1,\psi_3)$ for different values of parameter $C\in[0,1]$ (typical level lines for the surface~\eqref{eq:Cylinders}).
 }
\end{figure}

Three values of the disparity parameter stand out where the network structure is qualitatively different. 
As already discussed, $A=0$ corresponds to $K_\n=K_\s$, that is the strength of interactions within and between populations is the same. If in addition also $\alpha_\n=\alpha_\s$, the oscillators have full permutational symmetry and we have the Kuramoto--Sakaguchi equations for identical oscillators.
If $A=1$, we have $K_\n=0$ and the populations are uncoupled; this means that the green links are absent in Fig.~\ref{fig:B-M_Ph-Config}(a).
If $A=-1$, we have $K_\s=0$ and the oscillators within populations are uncoupled. This configuration is equivalent to a ring of oscillators with nearest-neighbor coupling as illustrated in Fig.~\ref{fig:B-M_Ph-Config}(a): If the purple links are absent, for a given oscillator the two oscillators of the other population are the direct neighbors.

\subsubsection{Reduced Dynamics and Symmetries}

The phase-shift symmetry of~\eqref{eq:goveqfull} means that the system dynamics is effectively three-dimensional. We exploit this by rewriting the system in terms of phase differences $\psi_{1} := \theta_{1,1}-\theta_{1,2}, \psi_2 := \theta_{1,2}-\theta_{2,1},\psi_3 := \theta_{2,1}-\theta_{2,2}$. This yields the reduced system
\begin{subequations}\label{eq:goveqn}
\begin{align}
\begin{split}
\dot\psi_1 &=
 -\frac{K_\s}{2} \cos(\alpha_\s)\sin(\psi_1)
 +\frac{K_\n}{4} \Big(
 \sin(\alpha_\n+\psi_2)\\&\qquad
 -\sin(\alpha_\n+\psi _1+\psi _2)
 +\sin(\alpha_\n+\psi _2+\psi _3)\\&\qquad
 -\sin(\alpha_\n+\psi_1+\psi_2+\psi_3)
\Big)\\
\end{split}\\
 \begin{split}
 \dot\psi_2 &=
  \frac{K_\s}{4}\big(\sin(\alpha_\s+\psi_3)-\sin(\alpha_\s-\psi _1)\big)\\&\qquad
  +\frac{K_\n}{4}\Big(
  \sin(\alpha_\n-\psi _1-\psi _2)\\&\qquad
  {-2}\cos(\alpha_\n)\sin(\psi_2)
  -\sin(\alpha _\n+\psi _2+\psi _3)\Big)\\
 \end{split}\\
 \begin{split}
\dot\psi_3 &=
 -\frac{K_\s}{2} \cos(\alpha_\s)\sin(\psi_3)
 +\frac{K_\n}{4} \Big(
 -\sin(\alpha_\n-\psi_2)\\&\qquad
 -\sin(\alpha_\n-\psi _1-\psi _2)
 +\sin(\alpha_\n-\psi _2-\psi _3)\\&\qquad
 +\sin(\alpha_\n-\psi_1-\psi_2-\psi_3)
\Big).\\
\end{split}
\end{align}
\end{subequations}

If the coupling is fully symmetric, i.e., $A=0$ ($K_\s=K_\n$) and $\alpha_\s=\alpha_\n$, then the phase space is organized by invariant planes~$\Pc_{nm}$, the canonical invariant region, and its images under the symmetry. In the reduced system~\eqref{eq:goveqn}, these invariant planes are given by $\psi_1=0$, $\psi_3=0$, $\psi_2=0$, $\psi_1+\psi_2=0$, $\psi_2+\psi_3=0$, $\psi_1+\psi_2+\psi_3=0$, which split~$\Tor^3$ into six dynamically invariant regions, the canonical invariant regions and its images. The geometry is illustrated in Fig.~\ref{fig:Cube_Psi-var}(a) where one invariant region is bounded by coordinate planes and a light grey plane $\psi_3=2\pi-\psi_1-\psi_2$. The intersection of all planes correspond to full phase synchrony. As defined above, the set of {incoherent phase configurations} are the points where the (global) Kuramoto order vanishes; in phase-difference coordinates these are given by
\[
\Ic =
\set{(\psi_1,\psi_2,\psi_3)}
{e^{\imath\psi_1}+e^{\imath(\psi_1+\psi_2)}+e^{\imath(\psi_1+\psi_2+\psi_3)}=-1}.
\]

For arbitrary parameters $\alpha_\s$, $\alpha_\n$, $A$, the network~\eqref{eq:goveqfull} has dihedral symmetry~$\Sym_2\wr \Sym_2\equiv \Di_4$. In phase differences~\eqref{eq:goveqn}, the generators of the symmetry action~\eqref{eq:SymGen} are
\begin{align}
\gamma_\mathrm{r}: (\psi_1,\psi_2,\psi_3) &\mapsto (-\psi_3,-\psi_1-\psi_2,\psi_1),\\
\gamma_\mathrm{m}: (\psi_1,\psi_2,\psi_3) &\mapsto (\psi_1,\psi_2+\psi_3,-\psi_3)
\end{align}
that correspond to the rotation and reflection mentioned above. 
Solutions transform in a nontrivial way along variable~$\psi_2$ that describes the state of the two populations relative to one another under~$\gamma_\mathrm{r}$, $\gamma_\mathrm{m}$. The system also has parameter symmetries
\begin{align}
\gamma^{(-)}&: (\psi_1, \psi_2, \psi_3,\alpha_\s,\alpha_\n) \mapsto 
-(\psi_1, \psi_2, \psi_3,\alpha_\s,\alpha_\n),\\
\gamma^{(\alpha_\n)}&:
(\psi_2;\alpha_\n)\mapsto (\psi_2+\pi;\alpha_\n+\pi),\\
\label{eq:gamma_A}
  \gamma^{(A,t)} &: (\psi_2; A, t)\mapsto
  \begin{cases}
  (\psi_2+\pi;1/A,t), & A>0 \\
  (\psi_2+\pi;1/A,-t), & A<0
  \end{cases}
\end{align}
where~\eqref{eq:gamma_A} is time-reversing for $A<0$.

\subsubsection{Equilibria, invariant subspaces, and linear stability}

The four oscillator network~\eqref{eq:goveqn} supports phase-synchronized (or coherent) solutions. Specifically, there are fixed points
\begin{align}\label{eq:invariant_subspaces}
\SSzero &= (0,0,0),\\
\SSpi   &= (0,\pi,0).
\end{align}
The first fixed point corresponds to the phase configuration where all oscillators are in-phase (the point of full isotropy where all phases are equal). The second corresponds to the configuration where both populations are in-phase but anti-phase with respect to each other.
Linear stability of the solutions~$\SSzero$ and~$\SSpi$ is determined by the eigenvalues
\begin{subequations}\label{eq:Eig_SS0}
\begin{align}
 \lambda_1 &= \mp {K_\n} \cos{\alpha_\n} \\
 \lambda_2 &= \mp {\frac{1}{2}}K_\n \cos{\alpha_\n} -{\frac{1}{2}}K_\s\cos{\alpha_\s}\\
 \lambda_3 &= \mp {\frac{1}{2}}K_\n \cos{\alpha_\n} -{\frac{1}{2}}K_\s\cos{\alpha_\s}.\
\end{align}
\end{subequations}
respectively; the signs are for~$\SSzero$ and~$\SSpi$, respectively.

The incoherent phase configurations in~$\Ic$ can be parametrized as three lines
\begin{subequations}
\begin{align}
\Ic_0 &:= \set{(\pi,\varphi,\pi)}{\varphi\in\Tor}\\
\Ic_1 &:= \set{(\varphi,\pi-\varphi, \varphi)}{\varphi\in\Tor}\\
\Ic_2 &:= \set{(\varphi,\pi,-\varphi)}{\varphi\in\Tor},
\end{align}
\end{subequations}
such that $\Ic = \Ic_0\cup\Ic_1\cup \Ic_2$.
The set~$\Ic_0$ are the phase configurations where each population is incoherent (that is, the phase difference of the two oscillators is~$\pi$), the sets $\Ic_1$ and $\Ic_2$ correspond to phase configurations where oscillators in distinct populations have a phase difference of~$\pi$.

The line~$\Ic_0$ is a continuum of equilibria. Linearizing the equations, we obtain that linear stability is determined by the eigenvalues
\begin{subequations}\label{eq:Eig_SSpi}
 \begin{align}
 \lambda_1 &= 0\\
 \lambda_2 &= {\frac{1}{2}}K_\s \cos \left(\alpha_\s\right) - \frac{|K_\n|}{{2}\sqrt{2}} \sqrt{\cos \left(2 \alpha _\n\right)+ \cos (2 \varphi)}\\
 \lambda_3 &= {\frac{1}{2}}K_\s \cos \left(\alpha _\s\right) + \frac{|K_\n|}{{2}\sqrt{2}} \sqrt{\cos \left(2 \alpha _\n\right)+ \cos (2 \varphi)}\
\end{align}
\end{subequations}
Writing $a=\frac{K_\s}{4}\cos\alpha_\s$, $b=\frac{K_\n}{4}\cos(\alpha_\n+\varphi)$, $c=\frac{K_\n}{4}\cos(\alpha_\n-\varphi)$ we have that
$\lambda_{2,3}=2(a\mp\sqrt{bc})$.
This yields the linear stability properties of the equilibria: Phase configurations in~$\Ic_0$ are linearly stable if $a<0$ and $bc<a^2$ or if $a<\min\{0,\ -\sqrt{bc} \}$. How the equilibria bifurcate depends on~$\varphi$: The eigenvalues~\eqref{eq:Eig_SSpi} are a complex conjugated pair if $\varphi\in(|\alpha_\n-\frac{\pi}{2}|-\pi, -|\alpha_\n-\frac{\pi}{2}|)\cup(|\alpha_\n-\frac{\pi}{2}|,\pi-|\alpha_\n-\frac{\pi}{2}|),$ which suggests a (degenerate) Hopf bifurcation for $K_\s\cos\alpha_\s=0$. For other~$\varphi$, the eigenvalues are real. In the limiting case of global coupling ($K_\s=K_\n:=1$, $\alpha_\s=\alpha_\n=\alpha$) the eigenvalues evaluate to
\begin{align*}
\lambda_{2,3}&=\frac{1}{2}\left(\cos\alpha\mp\sqrt{\cos^2\alpha-\sin^2\varphi}\right).
\end{align*}

Now consider the incoherent phase configurations~$\Ic_1$ and~$\Ic_2$. These are invariant lines of~\eqref{eq:goveqn} for any parameter values. The dynamics on each of these lines is given by
\[
\dot\varphi=\frac{1}{2}(K_\n\cos(\alpha_\n)-K_\s\cos(\alpha_\s))\sin(\varphi).
\]
Thus, $\Ic_1$ and~$\Ic_2$ are continua of nonisolated fixed points if $\alpha_\s=\alpha_\n=\pm\frac{\pi}{2}$ as in the traditional Kuramoto--Sakaguchi equations. Linear stability of these fixed points for $\alpha_\s=\alpha_\n=\alpha=\pm\frac{\pi}{2}$ is determined by the nontrivial eigenvalues
\[
\lambda_{2,3}=\pm i\frac{1}{2}\sqrt{K_\s K_\n}\sin(\varphi).
\]
This implies that the set of incoherent phase configurations~$\Ic=\Ic_0\cup\Ic_1\cup\Ic_2$ are surrounded by sets of periodic orbits when $\alpha=\pm\frac{\pi}{2}$ when $K_\n K_\s>0$ ($|A|<1$). If $\alpha_\s,\alpha_\n\approx\pm\frac{\pi}{2}$ the invariant lines persist and are surrounded by spiral trajectories.

Finally, we consider the set of \emph{two-cluster synchronized phase configurations~$\Sc$}, which are the points $\Sym_2\times \Sym_2$ isotropy in the globally coupled system: These are the points where there are two clusters of two oscillators each. In phase difference coordinates on~$\Tor^3$ for~\eqref{eq:goveqn}, we define
\begin{align*}
\Sc_0 &:= \set{(0,\varphi,0)\in\Tor^3}{\varphi\in\Tor},\\
\Sc_1 &:=\set{(\varphi,0,-\varphi)\in\Tor^3}{\varphi\in\Tor},\\
\Sc_2 &:=\set{(\varphi,-\varphi,\varphi)\in\Tor^3}{\varphi\in\Tor}.
\end{align*}
and have $\Sc = \Sc_0\cup\Sc_1\cup\Sc_2$. The set~$\Sc_0$ consists of the phase configurations where each of the populations is phase synchronized---this means that $\Sc_0 = \Pc_{1}\cap \Pc_{2} = \sset{\psi_1=0}\cap \sset{\psi_3=0}$; cf.~Fig.~\ref{fig:Cube_Psi-var}.
The sets~$\Sc_1, \Sc_2$ are phase configurations where two oscillators of distinct populations are phase-synchronized.
The set~$\Sc_0$ is dynamically invariant for any values of~$\alpha_\s$,~$\alpha_\n$, and~$A$ as fixed point sets of a subgroup of~$\Di_4$; the other two sets are only invariant for certain values of the coupling parameters as we will discuss below.
Now the dynamics on~$\Sc_0$ is are determined by
\[
\dot\varphi=-K_\n\cos(\alpha_\n)\sin(\varphi).
\]
For $\alpha_\n\ne\pm\frac{\pi}{2}$ there are exactly two fixed points on this line that correspond to~$\SSzero$ and~$\SSpi$, respectively. For $\alpha_\n=\pm\frac{\pi}{2}$, the set $\Sc_0$ is a continuum of fixed points. Linearization yields the eigenvalues
\begin{align*}
\lambda_1 &=-K_\n\cos(\alpha_\n)\cos(\varphi),\\
\lambda_{2,3} &=-\frac{1}{2}K_\n\cos(\alpha_\n\mp\varphi)-\frac{1}{2}K_\s\cos(\alpha_\s).
\end{align*}

Note that the invariant planes $\Pc_{1}=\sset{\psi_1=0}$ and $\Pc_{2}=\sset{\psi_3=0}$ bound the phase differences within each population. Thus, $\Omega_{11,12}=0$ and $\Omega_{21,22}=0$.
By contrast, the phase difference~$\psi_2$ between the two populations may not be bounded---this corresponds to weak chimeras. In the following sections we will elucidate the dynamical mechanisms that lead to such solutions.

\section{Gradient dynamics for odd coupling}
\label{sec:GradientDyn}

If the coupling function is a pure sine function ($\alpha_\s=\alpha_\n=0$), the system~\eqref{eq:goveqfull} is a gradient system\footnote{Note that the dynamics $\alpha_\s=\alpha_\n=\pi$ are identical up to a sign change.}. More precisely, with the potential
\begin{align*}
V(\theta)&
=-\frac{K_\s}{4}(\cos(\theta_{1,1}-\theta_{1,2})+\cos(\theta_{2,1}-\theta_{2,2}))
\\&\qquad
-\frac{K_\n}{4}\Big(\cos(\theta_{1,1}-\theta_{2,1})+\cos(\theta_{1,1}-\theta_{2,2})\\&\qquad\qquad\quad
+\cos(\theta_{1,2}-\theta_{2,1})+\cos(\theta_{1,2}-\theta_{2,2})\Big),
\end{align*}
equation~\eqref{eq:goveqfull} can be written as
\begin{align}\label{eq:GradDyn}
\dot\theta_{\sigma,k}=\omega-\frac{\partial}{\partial\theta_{\sigma,k}}V(\theta)
\end{align}
for $\sigma,k\in\sset{1,2}$.
The system has a parameter symmetry given by the action of~$\gamma^{(A)}$ as defined in~\eqref{eq:gamma_A}. This allows to restrict the parameter range to $A\in(-1, 1)$ and makes the bifurcation behavior of the system for the special parameter values  $A=\pm 1$ (corresponding to uncoupled populations and a ring topology, respectively) more transparent.
Note that for these parameter values~$\gamma^{(-)}$ is an additional $\mathbb{Z}_2$-symmetry.

\subsection{Equilibria and their stability}\label{sec:GradEq}

We first analyze the equilibria and their stability for $\alpha_\s=\alpha_\n=0$. According to~\eqref{eq:Eig_SS0}, the stability of the coherent equilibrium~$\SSzero$ is determined by the eigenvalues
$\lambda_1=(A-1)/2,$
$\lambda_{2,3}=-1/2.$
Thus, $\SSzero$ is an attractor for $A<1$ and a saddle equilibrium otherwise. Similarly, the stability of the equilibrium~$\SSpi$ is determined by
$\lambda_1=(1-A)/2,$ 
$\lambda_{2,3}=-A/2$.
Thus, $\SSpi$ is a source for $A\le 0$, a saddle for $A\in(0,1)$, and a sink for $A\ge 1$ (with neutral linear stability along~$\Ic_0$ for $A=1$).
Finally, the linearization of the vector field at the incoherent invariant line~$\Ic_0=\set{(\pi,\varphi,\pi)}{\varphi\in\Tor}$ has eigenvalues~\eqref{eq:Eig_SSpi}, which evaluate to
\begin{align*}
\lambda_1&=0, & \lambda_{2,3}&=\frac{1}{4}(1+A\pm|1-A||\cos(\varphi)|).
\end{align*}
Therefore, different points~$\Ic_0$ have different transverse stability for certain values of the parameter~$A$.
Specifically, for $A<-1$ the equilibria~$(\pi,\varphi,\pi)$ are stable if
\begin{align*}
\varphi&\in(-\varphi^*,\varphi^*-\pi)\cup(\pi-\varphi^*,\varphi^*),
\end{align*}
where $\varphi^*=\arccos\!\big(\frac{1+A}{|1-A|}\big).$
Decreasing~$A$ from zero, the equilibria on~$\Ic_0$ are (transversely) stable around $\varphi=\pm\frac{\pi}{2}$; the part of~$\Ic_0$ with transverse stability increases as~$A$ is decreased. Thus, there is multistability of the equilibrium~$\SSzero$ and segments of the manifold~$\Ic_0$ if $A<-1$. Conversely, for $A\in(-1,0)$ equilibria on $\Ic_0$ are repellers if $\varphi\in(\varphi^*-\pi,-\varphi^*)\cup(\varphi^*,\pi-\varphi^*)$.
For $A>0$, the entire line~$\Ic_0$ is repelling.
In all other cases, the points of $\Ic_0$ are of saddle type.

The system has a number of equilibria for apart from~$\SSzero, \SSpi$, and the continuum~$\Ic_0$. First, the points $(\pi,0,0)$, $(0,0,\pi)$, $(\pi,\pi,0)$, and $(0,\pi,\pi)$---two cluster configurations of one and three oscillators; cf.~Fig.~\ref{fig:B-M_Ph-Config}(k,l)---are equilibria for any~$A$ when $\alpha_\s=\alpha_\n=0$. Linear stability is determined by the eigenvalues
\begin{align*}
\lambda_1&=-\frac{K_\s}{2}, &
\lambda_{2,3}&=\frac{1}{4}\left(
K_\s\pm\sqrt{K_\s^2+8K_\n^2}
\right).
\end{align*}
These equilibria are of saddle type since
$K_\n>0$ implies $\lambda_1\lambda_2<0$,
$K_\n<0$ implies $\lambda_1\lambda_3<0$, and
$K_\n=0$ implies $\lambda_2\lambda_3<0$.

Second, for  $A>0$ the system has four symmetric equilibria $(-2\psi^*,\psi^*,0)$, $(2\psi^*,-\psi^*,0)$, $(0,\psi^*,-2\psi^*)$, and
$(0,-\psi^*,2\psi^*)$
with $\psi^*=\arccos\left(
\frac{A-1}{A+1}
\right)$. These equilibria are of saddle type and move along a straight line as the parameter~$A$ is varied. Pairs of these saddle points (in each plane $\psi_1=0$, $\psi_3=0$) disappear in pitchfork bifurcations with $\SSpi$ at $A=0$.

\subsection{Bifurcations and integrability}

The system bifurcates at $A=-1$, $A=0$, and $A=1$. In each of these cases, the dynamics have additional symmetries and there can be conserved quantities.

If $A=0$ the system is fully symmetric and the well-known Kuramoto model for identical oscillators. The system has a global attractor~$\SSzero$ and repellers $\Ic = \Ic_0\cup\Ic_1\cup\Ic_2$ consisting entirely of equilibria. Other equilibria with phase difference coordinates~$0$ and~$\pi$ are saddles in this case. breaking the full permutational symmetry for $A\ne 0$ leads to the disappearance of fixed points from invariant varieties~$\Ic_1$ and~$\Ic_2$ but these lines remain invariant.

For $A=\pm 1$ the system has conserved quantities.
If $A=1$, the oscillator populations are uncoupled and there is a conserved quantity. A straightforward calculation confirms:
\begin{prop}
For $A=1$ and any $\alpha_\s,\alpha_\n$ (thus also for $\alpha_\s=\alpha_\n=0$) the system has a first integral
\[
H^{(1,\,\cdot\,)}(\psi_1,\psi_3)=\cot\!\left(\frac{\psi_1}{2}\right)\tan\!\left(\frac{\psi_3}{2}\right).
\]
\end{prop}
\noindent Moreover for the specific gradient cases, we have an additional constant of motion; thus trajectories are completely determined.
\begin{prop}\label{prop:1,0}
For $A=1$ and $\alpha_\s=0$ the system has a first integral
\begin{align}
{H}^{(1,0)}(\psi_1,\psi_2,\psi_3)&=\psi_1+2\psi_2+\psi_3
\end{align}
\end{prop}
If $A=-1$ there is no coupling between distinct populations and we have a network of identical Kuramoto oscillators on a ring. This system also has a conserved quantity.
\begin{prop}
\label{prop:HmO}
For $A=-1$ and $\alpha_\s=\alpha_\n=0$ the system has a first integral
\begin{equation}\label{eq:HmO}
{H}^{(-1,0)}(\psi_1,\psi_3)=\cot\!\left(\frac{\psi_1+\psi_3}{4}\right)\tan\!\left(\frac{\psi_1-\psi_3}{4}\right)
\end{equation}
\end{prop}

\begin{figure}
\begin{center}
 \includegraphics[width=\linewidth]{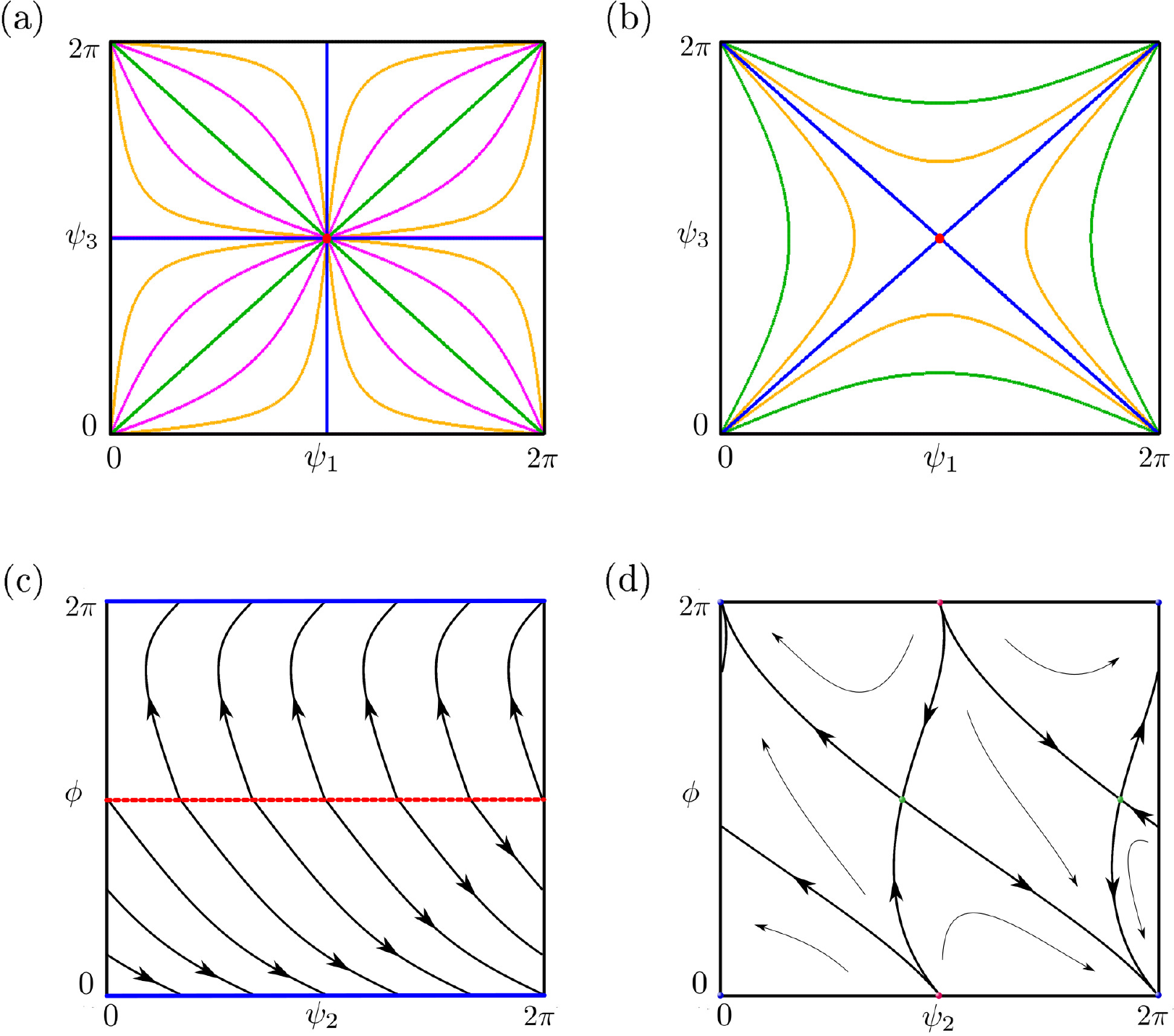}
 \end{center}
 \caption{\label{fig:First_Int_A}
Level lines of the first integrals $H^{(1,\,\cdot\,)}(\psi_1,\psi_3)=C$, $H^{(-1,\,0\,)}(\psi_1,\psi_3)=C$ for different constants $C$ (projection of system solutions on the plane $(\psi_1,\psi_3)\in\mathbb{T}^2$). (a) Level lines for the first integrals of the system (\ref{eq:goveqn}) for $A=1$ and arbitrary $\alpha_\s$; (b) the same for $A=-1$ and $\alpha_\s=\alpha_\n=0$. (c) and (d) show schematic phase portraits on level surfaces $H^{(1,\,\cdot\,)}=C$ and $H^{(-1,\,0\,)}=C$ in coordinates $(\psi_2,\phi)$, where $\phi$ is a parametrization variable along one of the level lines. The red dots in (a) and (b), as well as the red line in (c) correspond to the invariant manifold $\mathcal{I}_0$.
}
\end{figure}

Note that the constant of motion~$H^{(1,\cdot)}$, $H^{(-1,0)}$ do not depend on the inter population phase difference~$\psi_2$. Thus, all solutions belong to cylinder-like surfaces. Figure~\ref{fig:First_Int_A} shows the projections of surfaces determined by $H^{(1,\cdot)}(\psi_1,\psi_3)=C$ and $H^{(-1,0)}(\psi_1,\psi_3)=C$ for fixed $C\in [-1,1]$ onto the plane $(\psi_1,\psi_3)$ for different values of parameter~$C$.
In addition, for $A=1$ all solutions of the system belong to the parallel planes ${H}^{(1,0)}=\beta$, $\beta\in\Tor$: Each regular trajectory starts from a point of manifold~$\Ic_0$ and tends to a point of the set $\psi_1=\psi_3=0$; see Fig.~\ref{fig:First_Int_A}(c). The manifold~$\Sc$ consists completely of fixed points in this case.
For $A=-1$, the whole manifold~$\Ic$ consists of saddle equilibria (transverse to~$\Ic$).
In addition, the system has two more invariant lines of degenerate saddle points given by $\psi_1=\pi$, $2\psi_2+\psi_3=0$ and $\psi_1+2\psi_2=0$, $\psi_3=\pi$.
Stable and unstable 1-dimensional invariant manifolds of each saddle belong are contained in the same cylinder ${H}^{(-1,0)}=C$ for fixed~$C$. (Again, the saddles are neutral in the third direction).
All trajectories that are not equilibria or separatrices connect the source~$\SSpi$ to the attractor~$\SSzero$; see Fig.~\ref{fig:First_Int_A}(d). In particular, these trajectories on each cylinder ${H}^{(-1,0)}=C$ are bounded by 1-dimensional invariant manifolds of two saddle points.

To summarize, for the gradient case we observe the following stability/bifurcation behavior.
For $A<0$ the system has multistability of~$\SSzero$ and two parts of the manifold~$\Ic_0$ described in Sect.~\ref{sec:GradEq}. Moreover, the equilibrium~$\SSpi$ is a repeller.
For $A\in(-1,0)$, the system has only one attractor~$\SSzero$ and repellers~$\SSpi$ and parts of $\Ic_0$.
For $A>0$ the entire line $\Ic_0$ becomes a repeller.
For $A\in(0,1)$ the equilibrium $\SSzero$ is only attractor.
Finally, if $A>1$ the equilibrium~$\SSpi$ is an attractor and~$\SSzero$ is repelling.
One-dimensional manifolds of equilibria only exist at the bifurcation values of~$A$; except for $\Ic_0$ these do not persist as the degeneracies are broken.


\section{From conservative dynamics for even coupling to chimeras}
\label{sec:ConservativeDyn}

Weak chimeras as solutions where the two populations have distinct frequencies arise for phase-lag parameters close to~$\pm\frac{\pi}{2}$; cf. Refs.~\onlinecite{Ashwin2014a,Martens2016,Bick2017}. Here we elucidate the mechanisms that lead to such solutions. We mainly consider the singular case $\alpha_\s=\alpha_\n=\alpha=\pm\frac{\pi}{2}$: Then the vector field is 
{conservative (Hamiltonian-like)}. Indeed, if the coupling function is even (this is the case if $\alpha_\s=\alpha_\n=\pm\frac{\pi}{2}$), the system~\eqref{eq:goveqn} with vector field~$G$, is divergence-free, that is,
\[
\div(G):=\sum_{j=1}^{3}\frac{\partial G_j(\Psi)}{\partial\psi_j}\equiv 0
\]
for any values of~$K_\s$ and~$K_\n$; see also Ref.~\onlinecite{Ashwin2016}. This means that the system does not have any attractors or repellers and there are one-parameter families of fixed points and two-parametric families of periodic orbits or families of homo/heteroclinic cycles. Moreover, the system has the time-reversing symmetry given by action
\begin{align}
\grev &: (\psi_1, \psi_2, \psi_3, t) \mapsto (\psi_3, \psi_2, \psi_1, -t)
\end{align}
with fixed point space $\Fix(\grev) =  \set{(\psi_1,\psi_2,\psi_3)}{\psi_1=\psi_3}$
Due to the (parameter) symmetries, we can assume $\alpha:=\alpha_\n=\alpha_\s=\pm\frac{\pi}{2}$ without loss of generality. Indeed, for the other Hamiltonian-like case $\alpha_\s=-\alpha_\n=\pm\frac{\pi}{2}$ the parameter symmetries $\gamma^{(\alpha_\n)}$, $\gamma^{(A,t)}$ yield
\[
\gamma^{(\alpha_\n)}\circ\gamma^{(A,t)}: (A,\alpha_\s,\alpha_\n)\mapsto(A^{-1},\alpha_\s,\alpha_\n+\pi).
\]
Thus, the system~\eqref{eq:goveqn} with parameters $A$, $\alpha_\s=\pm\frac{\pi}{2}$, $\alpha_\s=\mp\frac{\pi}{2}$ is equivalent to same system with parameters $A^{-1}$, $\alpha_\s=\alpha_\n=\pm\frac{\pi}{2}$.

\subsection{Phase space and integrability}

Consider the case that the phase-lag parameter are identical, that is, $\alpha:=\alpha_\n=\alpha_\s=\pm\frac{\pi}{2}$. For $A=0$ ($K_\n=K_\s$) the system is fully symmetric and the phase space is organized into the canonical invariant region and its symmetric copies; cf.~Sect.~\ref{sec:2x2}. Specifically, the sets $\Sc_1$, $\Sc_2$ are dynamically invariant as points with isotropy $\Sym_2\times \Sym_2$. For $\alpha=\pm\frac{\pi}{2}$, these sets remain invariant even for nonidentical coupling strength $A\neq 0$ (this is not true in general): They form continua of equilibria whose linear stability is determined by the eigenvalues
\begin{subequations}
\begin{align}
\lambda_1&=0,\\
\lambda_{2,3}&=\pm\frac{1}{2}\sqrt{K_\s K_\n}\sin(\varphi).
\end{align}
\end{subequations}
The zero eigenvalue corresponds to the direction along~$\Sc_1$ or~$\Sc_2$, respectively, and the equilibria are degenerate saddles or degenerate centers depending on the sign $K_\s K_\n$ (or $1-A^2$).

By direct calculation, one can verify the existence of a first integral; see also the constants of motion in Refs.~\onlinecite{Panaggio2016,Watanabe1994}.

\begin{prop}
For $\alpha_\n=\alpha_\s=\pm\frac{\pi}{2}$ the system~\eqref{eq:goveqn} has the first integral
\begin{align*}
{H^{(\,\cdot\,,\frac{\pi}{2})}(\psi_1,\psi_3)}&=\sin\!\left(\frac{\psi_1}{2}\right)\sin\!\left(\frac{\psi_3}{2}\right).
\end{align*}
\end{prop}

The existence of the preserved quantity implies that the phase space is organized by invariant sets 
\begin{equation}\label{eq:Cylinders}
  \Lc(C)=\set{(\psi_1,\psi_2,\psi_3)}{{H^{(\,\cdot\,,\frac{\pi}{2})}}(\psi_1,\psi_3)=C, \psi_2\in\Tor}
\end{equation}
parametrized by $C\in(0,1)$. 
These are cylindrical 
if lifted to~$\mathbb{R}^3$ and two-dimensional tori in~$\Tor^3$; in the following, we will simply refer to~$\L(C)$ as \emph{invariant cylinders}. Note that we can also parametrize the invariant cylinders by their diameter~$d=4\sqrt{2}\arccos(\sqrt{C})$ (at $\psi_3=\psi_1$). One of these cylinders is shown in Fig.~\ref{fig:Cube_Psi-var}(c) and typical projections of such cylinders on the $(\psi_1,\psi_3)$-plane for different values of $C$ are shown in Fig.~\ref{fig:Cube_Psi-var}(d). In the limiting case of $C=1$, we have~$\Lc(C) = \Ic_0$. If $C=0$ it corresponds to the `square' cylinder of invariant planes $\psi_1=0$, $\psi_1=\pi$, $\psi_3=0$, $\psi_3=\pi$.
For $C\in(0,1)$, the set~$\Lc(C)$ contains eight equilibria, which correspond to the intersection of~$\Lc(C)$ with the sets $\Ic_1$, $\Ic_2$ and $\Sc_1$, $\Sc_2$:
With the parametrization of~$\Ic, \Sc$ as above, the intersection is at $\varphi=2\arcsin(\sqrt{C})$ and we have
\begin{align*}
\sset{I_1,I_1'} := \Ic_1\cap\Lc(C) &= \sset{(\varphi,\pi-\varphi,\varphi), (-\varphi,\varphi-\pi,-\varphi)},\\
\sset{I_2,I_2'} := \Ic_2\cap\Lc(C) &= \sset{(\varphi,\pi,-\varphi), (-\varphi,\pi,\varphi)},\\
\sset{S_1,S_1'} := \Sc_1\cap\Lc(C) &= \sset{(\varphi,0,-\varphi), (-\varphi,0,\varphi)},\\
\sset{S_2,S_2'} := \Sc_2\cap\Lc(C) &= \sset{(\varphi,-\varphi,\varphi), (-\varphi,\varphi,-\varphi)},
\end{align*}
The intersections are shown in Fig.~\ref{fig:Cube_Psi-var}(c): Each invariant line~$\Ic_\ell$, $\ell=1,2$, intersects the cylinder~$\Lc(C)$ twice, at the point~$I_\ell$ and the symmetric point $I_\ell'=-I_\ell$ (or $I_\ell'=2\pi-I_\ell$ in $\mathbb{R}^3$) for given $C\in(0,1)$. The invariant line $\Sc_\ell$ intersects the cylinder~$\Lc(C)$ at points~$S_\ell$ and its symmetric counterpart $S_\ell'=-S_\ell$.
In the limiting case $C=0$ the points $I_1$, $I_1'$, $I_2$, $I_2'$ correspond to~$\SSpi$ (and its symmetric copies) while $S_1$, $S_1'$, $S_2$, $S_2'$ correspond to the fully synchronized phase configuration~$\SSzero$.

\begin{figure}
\begin{center}
 \includegraphics[width=\linewidth,height=11cm]{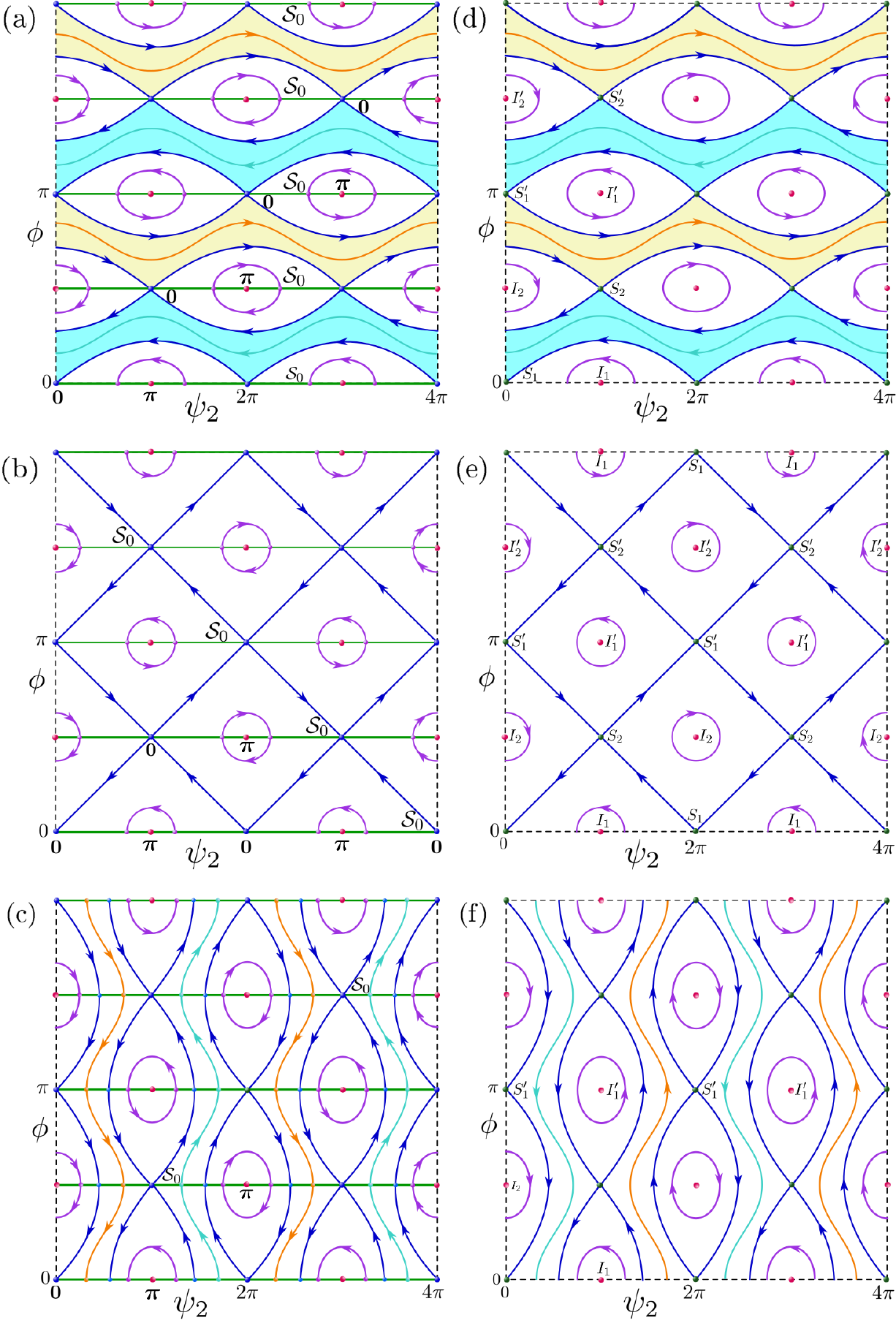}
 \end{center}
 \caption{\label{fig:Dyn_On_Surf}
 The phase portraits on the surface of the cylinder~$\Lc(C)$ for~$\alpha_\s=\alpha_\n=\pm\frac{\pi}{2}$ vary for different parameters $A\in[0,1)$ and~$C$. We show schematic phase diagrams on~$\Lc(C)$ in the variables $(\psi_2,\phi)$, where $\phi\in\Tor^1$ is an angle that parametrizes the curve $\sin(\psi_1/2)\sin(\psi_3/2)=C$. Phase portraits are skewed along the horizontal variable ($\psi_2\mapsto\psi_2+2\phi$) for a better presentation of symmetries of the original system~\eqref{eq:goveqfull}. 
The left row shows the dynamics on the boundary cylinder, which consists of invariant planes: $A\in (0,1)$, $C=0$ in Panel~(a); $A=0$, $C=0$ in Panel~(b); and $A\in (-1,0)$, $C=0$ in Panel~(c).
Green lines indicate the intersections of invariant planes~$\Pc_1, \Pc_2$.
The right row sketches the dynamics on the cylinders inside the invariant region: 
$A\in (0,1)$, $0<C<C^*=C^*(A)<1$ in Panel~(d); 
$A\in (0,1)$, $C=C^*\in(0,1)$ in Panel~(e); and
$A\in (-1,1)$, $1>C>C^*>0$ in Panel~(f).
Red points indicate the intersection of~$\Ic_\ell$ with~$\Lc(C)$, green points indicate intersections~$\Sc_\ell$ with~$\Lc(C)$ (they correspond to~$\SSpi$ in~(b)), blue points are~$\SSzero$. Light blue and yellow regions correspond to one-parametric families of periodic  chimeras that have opposite directions. 
Blue lines show homoclinic and heteroclinic connections, which bound the regions with phase unlocked dynamics in~(a) and~(d).
}
\end{figure}

\subsection{The emergence of chimera-like solutions}\label{sec:ChimDeg}

The global dynamics are determined by the dynamics on the invariant cylinders.
To obtain coordinates on the invariant cylinder (torus), write the variables $\psi_1$, $\psi_3$ in polar coordinates~$\rho$, $\phi$. Now the two-dimensional dynamics on~$\Lc(C)$ can be expressed in  coordinates~$(\psi_2,\phi)$; these depend on the value of~$C$ and may bifurcate as~$C$ is varied. Note that the dynamics for any~$C$ is present since~$C$ parametrizes the family of invariant tori that foliates phase space rather than being a system parameter~\cite{Liebscher2011,Fiedler2000}.
First consider the special case $C=0$; in this case~$\Lc(C)=\Lc(0)$ corresponds to the degenerate cylinder that consists of the invariant planes. The dynamics on~$\Lc(0)$ depend on the coupling parameter~$A$ as shown in Fig.~\ref{fig:Dyn_On_Surf}(a--c): For $A\in(0,1)$ there are continua of heteroclinic trajectories from equilibria in~$\Sc_0$ as shown in Fig.~\ref{fig:Dyn_On_Surf}(a); these are bounded by homoclinic trajectories from~$\SSzero$ to itself (on the torus) and degenerate to the equilibrium~$\SSpi$. Moreover, there are families of periodic orbits with nontrivial winding number (shaded areas).
For~$A=0$ these families of periodic orbits disappear as the stable and unstable manifold of~$\SSzero$ form a heteroclinic ``web'' on~$\Lc(0)$; see Fig.~\ref{fig:Dyn_On_Surf}(b). Finally for $A\in(-1,0)$, there are homoclinic/heteroclinic trajectories from (points in)~$\Sc_0$ to itself.
For fixed $A\in(0,1)$ the bifurcation scenario is similar with~$\SSzero$ replaced by $S_j,S_j'$, $j=1,2$ and~$\SSpi$ replaced by $I_j,I_j'$, $j=1,2$; see Fig.~\ref{fig:Dyn_On_Surf}(d--f).
Define the critical cylinder size
\begin{equation}
C^*=\begin{cases}
\displaystyle
\frac{2A}{A+1}, & A\in[0,1],\\[4mm]
\displaystyle
\frac{2}{A+1}, & A\ge1.
\end{cases}
\end{equation}
(Here, we used the symmetry $\gamma^{(A,t)}$ to obtain the formula for $A\ge 1$.)
For $C\in(0,C^*)$ we have families of periodic orbits where the populations rotate relative to one another. For $C = C^*$ we have a heteroclinic web between $S_j,S_j'$. Finally, for $C\in (C^*,1)$, there are periodic frequency locked solutions.

\subsection{No coupling between populations}
If $A=1$ ($K_\n=0$) the network consists of two uncoupled populations that evolve independently of one another. In this case, $\gamma^{(A,t)}$ acts as a symmetry of the system since it maps $A\mapsto A^{-1}$. The system~\eqref{eq:goveqn} reduces to
\[
(\dot\psi_1, \dot\psi_2, \dot\psi_3) = (0,\cos(\psi_3)-\cos(\psi_1),0)
\]
and its solutions form a continuum of lines
\[
(\psi_1,\psi_2,\psi_3)(t)=(\psi_1^0,(\cos(\psi_3^0)-\cos(\psi_1^0))t+\psi_2^0,\psi_3^0)
\]
parametrized by the initial conditions $(\psi_1(0),\psi_2(0),\psi_3(0))=(\psi_1^0,\psi_2^0,\psi_3^0)$. The dynamics are schematically shown in Fig.~\ref{fig:Cube_Psi-var}(b).
For most initial conditions, the phase difference~$\psi_2$ between the populations is increasing or decreasing monotonously. Specifically, we have that each population is frequency synchronized and they rotate relative to one another with $\Omega_{2,j}-\Omega_{1,k}=\cos(\psi_3(0))-\cos(\psi_1(0))$, $j,k=1,2$.
Thus, $\psi_2$ increases monotonically if $|\psi_3|>|\psi_1|$ for $\psi_1,\psi_3\in [-\pi,\pi]$ and $\psi_2$ decreases monotonically if $|\psi_3|<|\psi_1|$ for $\psi_1,\psi_3\in [-\pi,\pi]$.
The direction of the flow is shown in Fig.~\ref{fig:Cube_Psi-var}(b) in the cube $[0,2\pi]^3$.
On the planes $\psi_3=\pm\psi_1$ the dynamics are trivial (there are nonisolated fixed points). Note that the invariant lines~$\Ic_0$ and~$\Sc_0$ are the intersection of these planes.

\begin{figure}
\begin{center}
 \includegraphics[width=\linewidth]{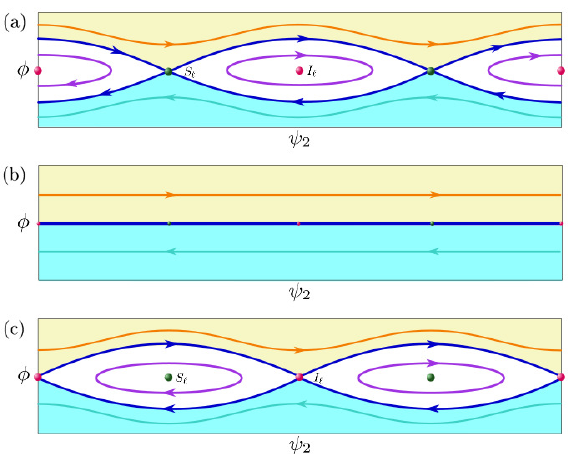}
 \end{center}
 \caption{\label{fig:Surf_Bif_A_1}
 There is a degenerate bifurcation on the cylinders~$\Lc(C)$, $C\in(0,1)$, (see~\eqref{eq:Cylinders}) at $A=1$ for $\alpha_\s=\alpha_\n=\pm\frac{\pi}{2}$ on.
 Panel~(a) gives a schematic for $A<1$ before the bifurcation; Panel~(b) at the bifurcation $A=1$; and Panel~(c) for $A>1$. 
 Phase portraits are given for $\psi_2\in [0,4\pi]$ and $\phi\in[-\phi^*,\phi^*]$ for $\phi^*$ small. Red and blue points (balls) indicate~$I_\ell$ and~$S_\ell$, respectively for $\ell=1,2$, blue lines are heteroclinic trajectories, light blue and yellow regions indicate continuous families of phase-unlocked (chimera) solutions. 
 The heteroclinic cycle connecting~$S_\ell$ is compressed to the line $\phi=0$ and a new heteroclinic cycle appears connecting~$I_\ell$. At the bifurcation point, the whole line $\phi=0$ consists of fixed points as~saddles~$S_\ell$ become centers and centers~$I_\ell$ become saddles. The points~$I_\ell'$ and~$S_\ell'$ undergo an analogous bifurcation elsewhere on~$\Lc(C)$.
}
\end{figure}

For coupling parameters~$A$ ($K_\n$) close to the singular case of uncoupled populations, the coupling between populations is weak. Frequency unlocked solutions persist but now deviate from straight lines. This gives the frequency unlocked solutions on the invariant cylinders $\Lc(C)$ for $C<C^*=C^*(A)$ and given $A\in(0,1)$ described above. For $\abs{A}>1$, the parameter symmetry $\gamma^{(A,t)}$ implies that the system~\eqref{eq:goveqn} has the same phase portraits but with the shift $\phi_2\mapsto\phi_2+\pi$ and a potential time reversal. The symmetry also means that the saddles~$S_\ell$, $S_\ell'$ swap places with centers~$I_\ell$, $I_\ell'$. Local bifurcations occur at each fixed point of invariant manifold of the system (two eigenvalues of each point transform from pure imaginary to real symmetric or vice versa). More globally all invariant manifolds of the fixed points~$\Sc_\ell$ swap with those of~$\Ic_\ell$, $\ell=1,2$.
Fig.~\ref{fig:Surf_Bif_A_1} schematically shows the bifurcation that occurs in the narrow stripe of phase space $(\psi_2,\phi)\in \Lc(C)$ at the bifurcation point $A=1$:
There is a heteroclinic cycle between the saddle~$S_\ell$ ($S_\ell'$) and its copy with coordinate $\psi_2+2\pi$ which bounds a family of periodic orbits around the center~$I_\ell$ ($I_\ell'$) (see~Fig.~\ref{fig:Surf_Bif_A_1}(a)). At the bifurcation point $A=1$, the phase space is foliated by straight lines as discussed above (see Fig.~\ref{fig:Surf_Bif_A_1}(b)).
For $A>1$ a new heteroclinic loop appears between the saddle~$I_\ell$ ($I_\ell'$) that bounds periodic orbits centered at~$S_\ell$ ($S_\ell'$).
Frequency unlocked chimera solutions exist for all $A>1$.

\begin{figure}
\begin{center}
 \includegraphics[width=\linewidth]{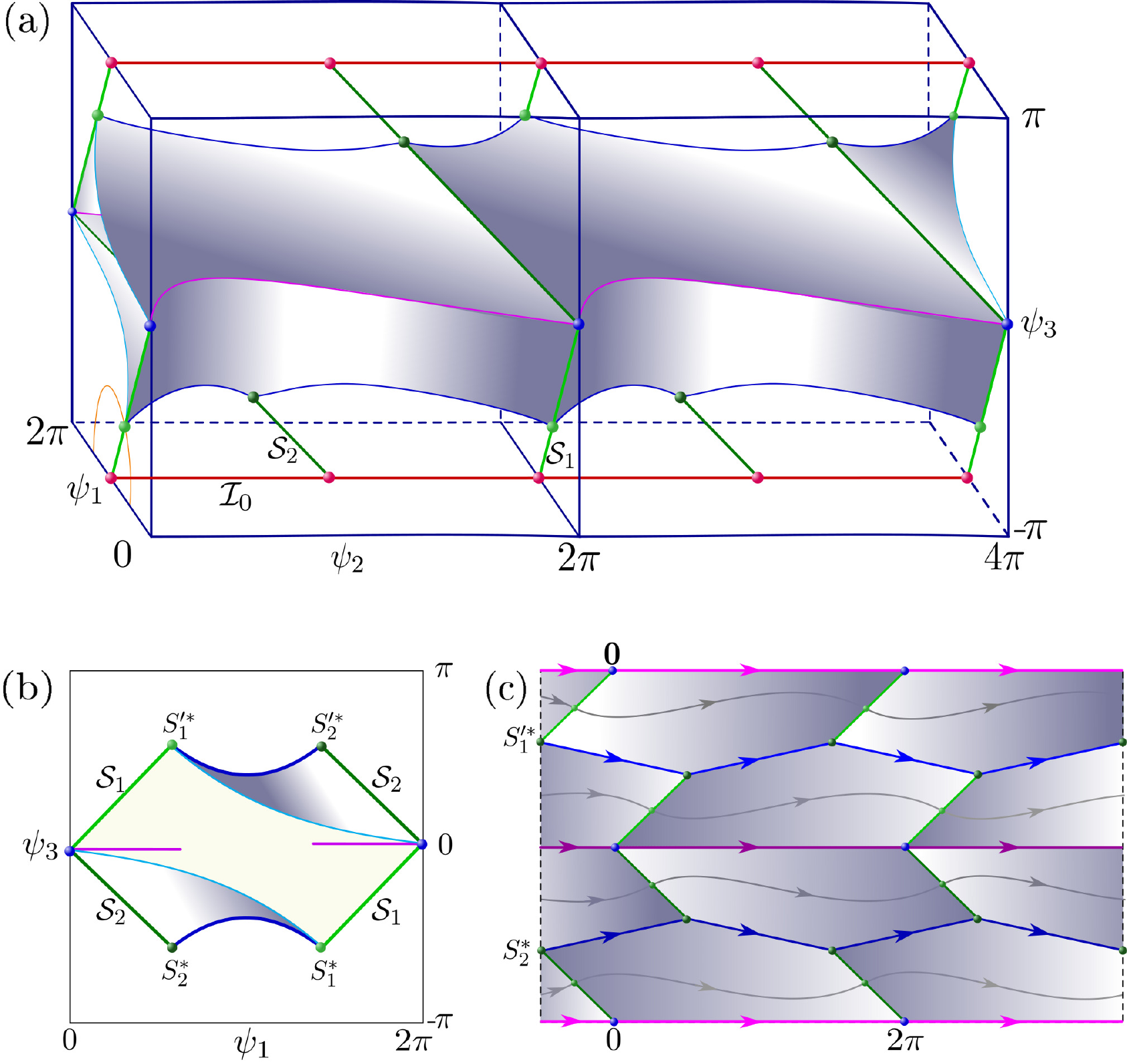}
 \end{center}
 \caption{\label{fig:Chimera-Snake}
 Phase unlocked dynamics between populations exist with a serpentine chimera region for $(\psi_1,\psi_2,\psi_3)\in [0,2\pi]\times[0,4\pi]\times[-\pi,\pi]$. 
 Panel~(a) shows the boundary surface (grey shading) in~$\Tor^3$ of the region. It consists of a one-parametric family of homoclinic trajectories between the points~$S_\ell$, $S_\ell'$ (green points and lines), two heteroclinic trajectories of points $S_\ell^*:=S_\ell(C^*)_2$ and $S_\ell'^*:=S_\ell'(C^*)$ (blue lines)---on the cylinder~$\Lc(C^*)$---and two homoclinic cycles of points~$\SSzero$ (magenta lines). The surface consists of four symmetric smooth pentagonal parts, some of which are related by symmetry. Red lines indicate~$\Ic_0$; green lines indicate~$\Sc_1$, $\Sc_2$; the magenta line is a homoclinic orbit of~$\SSzero$ within $\Pc_2=\sset{\psi_3=0}$; the orange semicircle is a projection of~$\Lc(C)$ onto $\psi_2=0$. 
 Panel~(b) shows a projection of the region onto $\psi_2=0$. 
 Panel~(c) shows the dynamics on the surface of the region flattened out; the magenta/purple lines are related by symmetry.
}
\end{figure}

We can now describe the fate of the frequency-unlocked chimera solutions in more detail. At the bifurcation point $A=1$ the whole phase space is filled with frequency unlocked chimera solutions except on the invariant planes $\psi_3=\pm\psi_1$. Indeed, there are two `chimera tubes' with opposite directions of trajectories.
As the parameter~$A$ deviates from one, narrow cylinders with phase-locked periodic orbits appear. The heteroclinic trajectories form the boundaries between the region of frequency locking and a region of frequency-unlocked chimera solutions. Tracing these trajectories along the invariant cylinders $\Lc(C)$ that foliate phase space for $C\in (0,C^*)$, we obtain a tube of phase unlocked chimera solutions; cf~Fig.~\ref{fig:Chimera-Snake}. Since the surface bounding the tube resembles a snake, we will refer to it as a \emph{serpentine chimera region}. Fig.~\ref{fig:Chimera-Snake}(c) shows boundary surface of the tube schematically (``skin of the chimera-snake'') that consists of piecewise smooth surfaces. As $A\to 0$ the tube degenerates to a single heteroclinic connection with nontrivial winding number. Given the parameter symmetry~$\gamma^{(A,t)}$, we can conclude that there are phase unlocked chimera solutions for any $A>0$.

\subsection{No coupling within populations}

Finally, consider the case $A=-1$ when the oscillators within each population are uncoupled ($K_\s=0$)---they still remain coupled indirectly through the interaction with the other population. In this case, $\gamma^{(A,t)}$ acts as a time-reversing symmetry since it maps $A\mapsto A^{-1}$. A direct calculation shows:

\begin{prop}
For $A=-1$ and $\alpha_\n=\pm\frac{\pi}{2}$ the system~\eqref{eq:goveqn} has the conserved quantity
\[
H^{(-1,\frac{\pi}{2})}(\psi_1,\psi_2,\psi_3)=\psi_1+2\psi_2+\psi_3.
\]
\end{prop}

\begin{rem}
Note that $H^{(-1,\frac{\pi}{2})} = {H}^{(1,0)}$ in Proposition~\ref{prop:1,0}.
\end{rem}

Thus for $A=-1$, there are two conserved quantities, $H^{(\,\cdot\,,\frac{\pi}{2})}$ and $H^{(-1,\frac{\pi}{2})}$, and thus solutions lie on the intersection of the invariant cylinders~$\L(C)$ with the plane $H^{(-1)}=C^{(-1)}$ for~$C\in(0,1)$, $C^{(-1)}\in\Tor$, given by 
\[
\mathcal{K}(C,C^{(-1)})=\set{(\psi_1,\psi_2,\psi_3)}{\begin{array}{c}\sin\!\left(\frac{\psi_1}{2}\right)\sin\!\left(\frac{\psi_3}{2}\right)=C,\\ \psi_1+2\psi_2+\psi_3=C^{(-1)}\end{array}}.
\]
Similar to the case $A=1$, local and global bifurcations occur when $A=-1$.
The time reversing symmetry organizes the solutions on the invariant cylinders~\eqref{eq:Cylinders}: Invariant lines of saddle fixed points~$\Sc_\ell$ mutually change the structure of their fixed points with the lines of center fixed points~$\Ic_\ell$, $\ell=1,2$.
Global bifurcation with phase-locked periodic orbits (lines in phase space $(\psi_2, \phi)\in\mathbb{R}^2$) for $A=-1$ is of the same type that the global bifurcation with the appearance of the family of straight lines in the described case $A=1$. In the case $A=-1$ global bifurcations occur on the square $\mathcal{K}(0,0)\in\Tor^3$ (with vertices at the points $\SSzero$, $\SSpi$ and two their copies) instead of~$\Sc_0$ as for $A=1$.

\section{Dissipative dynamics: Bifurcations and transitions to chaos}
\label{sec:DissipativeDyn}

In the conservative case, limit cycle solution with nontrivial winding number along the phase difference~$\psi_2$ (the phase difference between the two populations) arise in phase space. How do these solutions bifurcate as parameters are varied and the additional structures are broken? In the following, we will refer to any limit cycle solution with nontrivial winding number as a weak chimera.

\subsection{From conservative to dissipative dynamics}

Before we turn to chimeras, we consider the overall organization of phase space by invariant subsets and qualitatively describe what dynamics are possible. The conservative dynamics for $\alpha_\s=\pm\alpha_\n=\pm\frac{\pi}{2}$ correspond to a bifurcation point. Recall that for these parameter values, the sets~$\Ic_1$, $\Ic_2$ and $\Sc_1$, $\Sc_2$ are continua of equilibria; in particular, the linearization at each equilibrium has a zero eigenvalue with eigenvector in the direction of the corresponding set. Away from the bifurcation point, there are only two equilibria for each set: These are
$\sset{\SSpi, (\pi,0,\pi)}\subset\Ic_1$,
$\sset{\SSpi, (\pi,\pi,\pi)}\subset\Ic_2$,
$\sset{\SSzero, (\pi,0,\pi)}\subset\Sc_1$, and
$\sset{\SSzero, (\pi,\pi,\pi)}\subset\Sc_2$.
Before the bifurcation point, one of the two equilibria is repelling along the set while the other one is attracting. After the bifurcation, the situation is reversed. Hence, the bifurcation corresponds to reversing the flow along the four sets in a degenerate bifurcation. The stability transverse to the invariant set determines the dynamics of trajectories nearby: If there are complex conjugate eigenvalues, nearby trajectories spiral around the set as the conservative structure is broken.

Away from the bifurcation point $\alpha_{\s},\alpha_\n=\pm\frac{\pi}{2}$, the invariant cylinders (Fig.~\ref{fig:Cube_Psi-var}(c)) break up which allows for more intricate dynamics. For small deviations from the bifurcation parameter $\alpha_\s=\pm\frac{\pi}{2}-\varepsilon$, there is a slow drift transverse to the cylinders~$\Lc(C)$ (which are not invariant anymore). Since the families of heteroclinic orbits that separate frequency-locked and frequency-unlocked solutions also disconnect, it is possible for trajectories to come close to the continuum of equilibria~$\Ic_0$ as well as the degenerate cylinder~$\Lc(0)$ that consists of the invariant planes $\psi_1=0$ and $\psi_3=0$.

Taken together, typical trajectories can exhibit dynamics that explore a large region of phase space beyond the bifurcation point.
Trajectories can move from the vicinity of the points~$\SSpi$~($\SSzero$), moving along the invariant line~$\Ic_1$ or~$\Ic_2$ ($\Sc_1$ or~$\Sc_2$) in a spiraling fashion to approach~$\Ic_0$, before then drifting along the level sets~$\Lc(C)$ into the region where the phase difference between populations increases and approaching a frequency-unlocked solution on the invariant set~$\Lc(0)$. In the following we will consider the bifurcations of these asymptotic solutions.

\subsection{Bifurcations on invariant manifolds and flat chimeras}

We first consider the weak chimeras on the invariant planes~$\Pc_1, \Pc_2$ defined by $\psi_1=0$ or $\psi_3=0$ (these correspond to the limiting cylinder~$\Lc(0)$ in Sect.~\ref{sec:ChimDeg}); we will refer to these solutions as \emph{flat chimeras}. These correspond to in-phase synchronization of one population while the other population is approximately in anti-phase~\cite{Ashwin2014a,Bick2017}. The relative phase difference between the two populations increases without bound as $\Omega_{1,2}\ne 0$. Without loss of generality, we assume that $\psi_3=0$, that is, the second population is phase synchronized. The dynamics~\eqref{eq:goveqn} on the invariant subspace restrict to a two dimensional system; therefore no chaotic flat chimeras are possible.
Recall that families of flat chimeras arise for $\alpha_\s=\pm\frac{\pi}{2}$, $\alpha_\n=\pm\frac{\pi}{2}$ as described above; cf.~Fig.~\ref{fig:Dyn_On_Surf}(a). This family of flat chimeras is usually bound{ed} by homoclinic or heteroclinic cycles (Fig.~\ref{fig:Dyn_On_Surf}(b)). These flat chimeras are neurally stable transverse to the invariant plane~$\psi_3=0$ due to the (partial) integrability of the system.

\makeatletter\onecolumngrid@push\makeatother
\begin{figure*}
\begin{center}
 \includegraphics[width=1\linewidth,height=8cm]{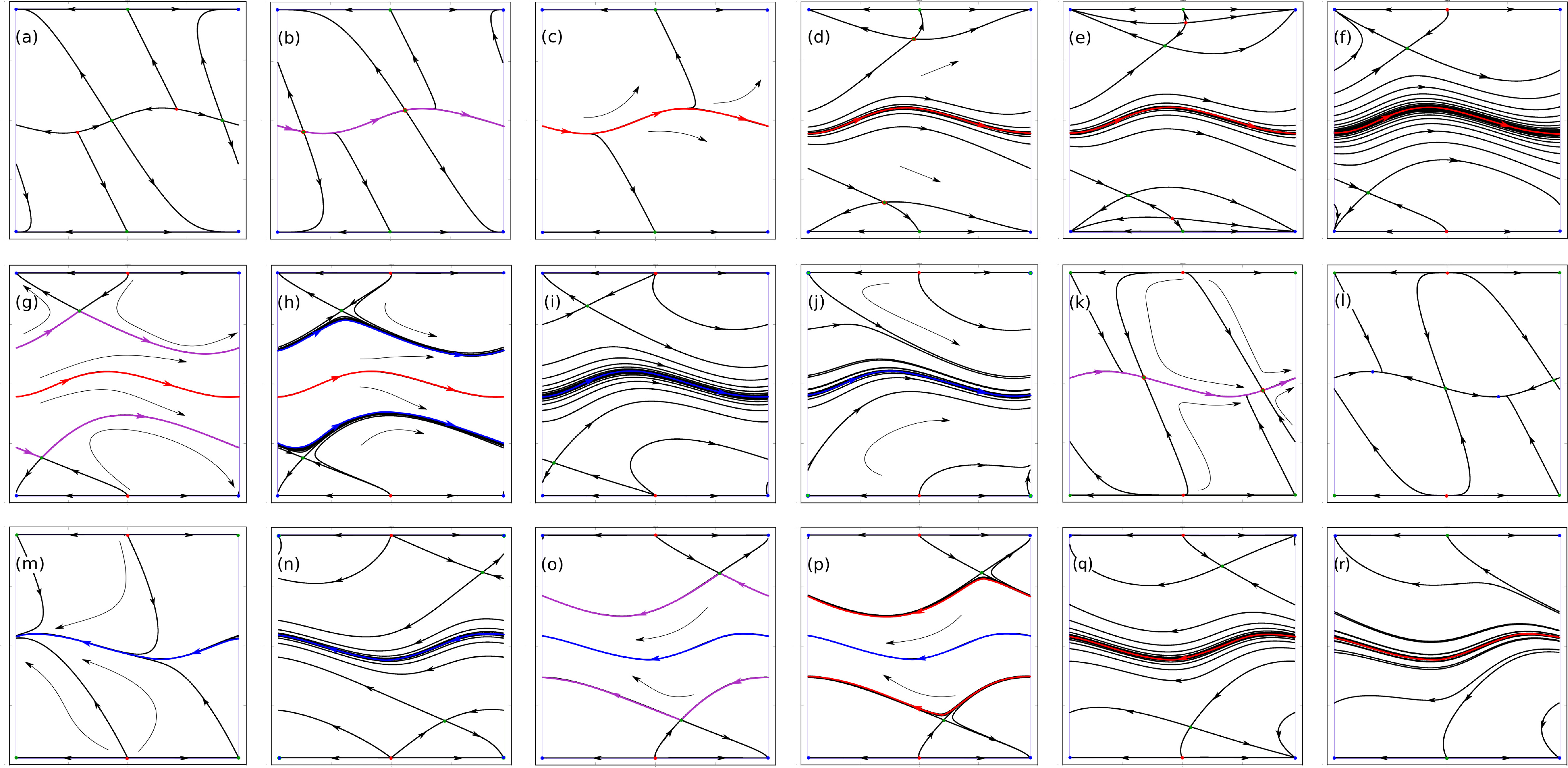}
 \end{center}
 \caption{\label{fig:P3_0}
 Phase portraits on the invariant surface~$\Pc_2 = \sset{\psi_3=0}$ for $(\psi_2,\psi_1)\in[0,2\pi]^2$ show the bifurcation behavior for increasing parameter $\alpha_\s\in[0,2\pi]$ keeping $A=0.7$, $\alpha_\n=0.44$ fixed. Specifically, we have
 Panel~(a) $\alpha_\s = 0$;
 (b) $\alpha_\s = 0.041143$;
 (c) $\alpha_\s = 0.1$;
 (d) $\alpha_\s = 1.378599$;
 (e) $\alpha_\s = 1.4$;
 (f) $\alpha_\s = 1.410375$;
 (g) $\alpha_\s = 1.55$;
 (h) $\alpha_\s = 1.64$;
 (i) $\alpha_\s = 1.731144$;
 (i) $\alpha_\s = 3.16$;
 (k) $\alpha_\s = 3.5$;
 (l) $\alpha_\s = 4.520192$;
 (m) $\alpha_\s = 4.687983$;
 (n) $\alpha_\s = 4.69$;
 (o) $\alpha_\s = 4.8$;
 (p) $\alpha_\s = 4.9$.
 Fixed points are colored according to type: source (red), sink (blue), saddle (green), saddle-node (two-color). 
 Stable limit cycles are shown in blue, unstable in red, homo/heteroclinic in magenta. All cycles are non-homologic to zero on the two-dimensional torus and they correspond to weak chimeras.
}
\end{figure*}
\clearpage
\makeatletter\onecolumngrid@pop\makeatother

Fig.~\ref{fig:P3_0} elucidates the bifurcation structure for parameter values for parameter values where the conservative structure is broken.
Note that because of the symmetries, the system can have only {an} even number of fixed points on the plane $\psi_3=0$. The system has 2, 4, or 6 fixed points depending on parameters values. Saddle-node and saddle-connection bifurcations must occur in pairs simultaneously.
Specifically, the panels in Fig.~\ref{fig:P3_0}(a--p) show all possible phase portraits for the dynamics of $(\psi_2,\psi_1)\in\Tor^2$; these are arranged to show bifurcation transitions from~(a) to~(b), from (b)~to~(c), and so on.
The system {exhibits} the following bifurcation{s} on the plane: A pitchfork bifurcation of~$\SSzero$ or~$\SSpi$ transverse to the invariant line $\psi_1=0$ (Fig.~\ref{fig:P3_0}(c--d), (k--l), (l--m), (o--p)); two simultaneous saddle-node bifurcations on the flat chimera limit cycle (Fig.~\ref{fig:P3_0}(a--c), (i--k)); two simultaneous saddle-node bifurcations away from the chimera (Fig.~\ref{fig:P3_0}(o--n)); a subcritical pitchfork bifurcation of limit cycles (Fig.~\ref{fig:P3_0}(h--g)); a supercritical pitchfork bifurcation of limit cycles (Fig.~\ref{fig:P3_0}(g--h))); two simultaneous saddle connection bifurcations (Fig.~\ref{fig:P3_0}(f--g), (m--n)).

We highlight the bifurcations that relate to the creation and destruction of stable flat chimeras. First, there are two simultaneous saddle-node bifurcations on an invariant circle that lead to a limit cycle solution with nonzero winding number, a flat chimera (Fig.~\ref{fig:P3_0}(a--c), (k--m)). 
Second, {there} are pitchfork bifurcation of limit cycles both sub- (Fig.~\ref{fig:P3_0}(h,i)) and supercritical (Fig.~\ref{fig:P3_0}(p,q)); these bifurcations can stabilize flat chimeras (within the invariant subspace).
Third, there are simultaneous saddle-connection bifurcations of saddle equilibria that lead to the emergence of two symmetric limit cycles with unbounded phase difference~$\psi_2$ between populations; the resulting limit cycles can be stable (Fig.~\ref{fig:P3_0}(f--g)) or unstable (Fig.~\ref{fig:P3_0}(m--n)).

Note that the stability of flat chimeras in the full system~\eqref{eq:goveqn} depends on the transverse stability with respect to the third direction.


\subsection{Chaotic weak chimeras}\label{sec:chaoticweakchimeras}

Two populations of two phase oscillators with Kuramoto--Sakaguchi coupling support chaotic weak chimeras; the bifurcation scenario, obtained numerically\footnote{The bifurcation diagram was obtained via quasi-continuation, i.e., the dynamical equations were numerically integrated over $T=10000$ time units for each parameter value, whereas end solutions for each parameter value were used as initial condition for the next parameter value.
The initial condition $(\psi_1,\psi_2,\psi_3)=(4.7204,1.6028,3.1790)$ was used as starting point at $\alpha_s=1.6$ for quasi-continuation with endvalues $\alpha_s=1.56$ and $\alpha_s=1.6415$ in backward and forward directions, respectively.}, is indicated in Fig.~\ref{fig:BifDiagChaos}.

In the following we describe the bifurcations that lead to the emergence of chaotic weak chimeras in more detail. Since $\abs{\Di_4}=8$, every point with trivial isotropy has eight images under the symmetry action. Recall that~$\Sigma(A)$ are the symmetries that preserve the set~$A$. If~$A_\eta$ is a family of limit sets indexed by a parameter~$\eta$ and~$\Sigma(A_\eta)$ changes as~$\eta$ is varied, we have a \emph{symmetry increasing bifurcation}~\cite{Chossat1988}.

\makeatletter\onecolumngrid@push\makeatother
\begin{figure*}
\begin{center}
 \includegraphics[width=1\textwidth]{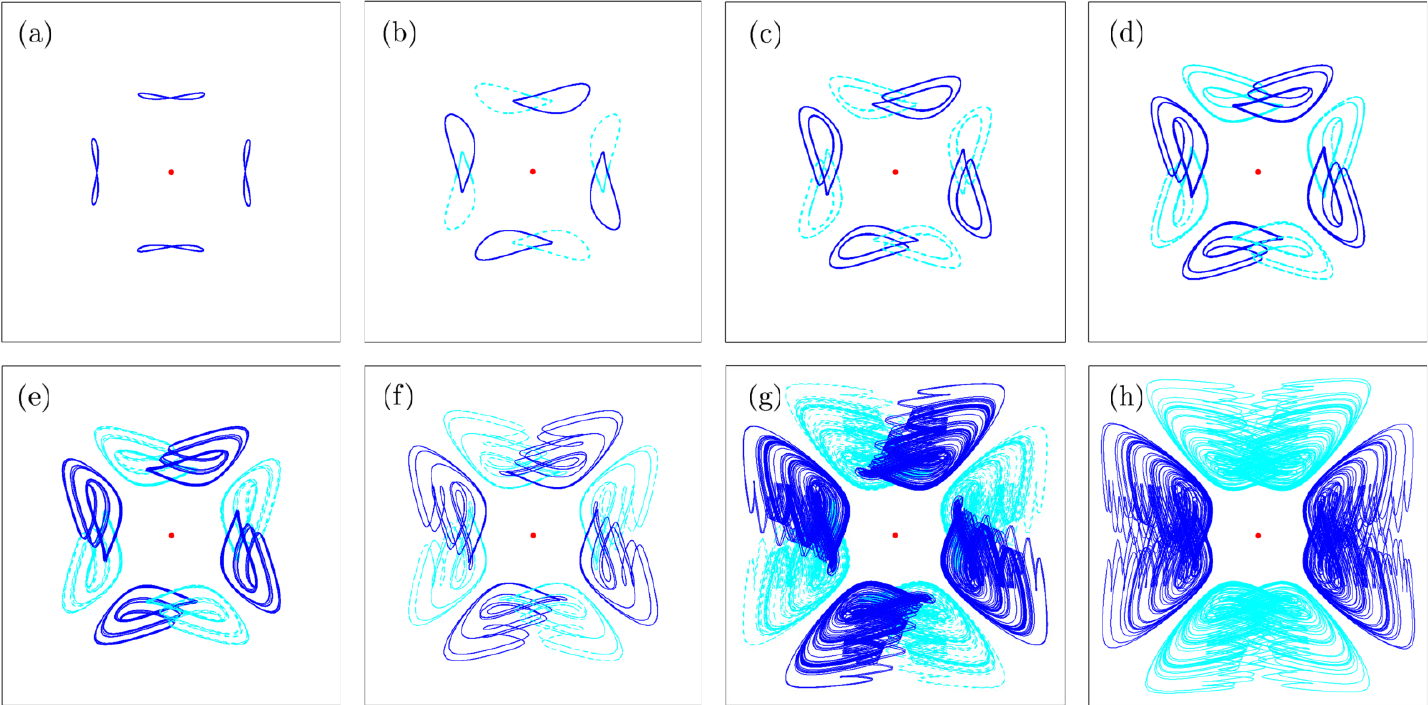}
 \end{center}
 \caption{\label{fig:B-M_Chaos-Proj}
 In the full three-dimensional phase space, the chimera attractors bifurcate and eventually become chaotic; here projections of the trajectories onto $(\psi_1,\psi_3)\in[0,2\pi]^2$ are shown where the~$\Di_4$ symmetry acts in the obvious way. Parameters are $A=0.7$, $\alpha_\n=0.44$ in all panels and~$\alpha_\s$ as in Fig.~\ref{fig:BifDiagChaos}.
 Panel~(a) shows self-symmetric eight-shape stable limit cycles
 for $\alpha_\s=1.607$;
   Panel~(b) shows pairs of stable limit cycles for $\alpha_\s=1.63$ that emerge from eight-shaped ones in a symmetry-breaking bifurcation;
  Panel~(c--f) show limit cycles after period-doubling bifurcations
 with $\alpha_\s=1.635$ in~(c), $\alpha_\s=1.638$ in~(d); $\alpha_\s=1.6385$ in~(e) and $\alpha_\s=1.639$ in~(f);
  Panel~(g) shows eight symmetric chaotic attractors
 for $\alpha_\s=1.6415$;
  Panel~(h) shows four symmetric chaotic attractors for $\alpha_\s=1.64166$ after a symmetry-increasing bifurcation.
 Line thickness was reduced in (f--h) for visualization. Line styles/colors are used to distinguish solutions that are related by symmetry. The red dot indicates~$\Ic_0$.
 }
\end{figure*}
\clearpage
\makeatletter\onecolumngrid@pop\makeatother

We describe the bifurcations as a single parameter is varied; for concreteness, we fix $A=0.7$, $\alpha_\n=0.44$ and vary as~$\alpha_\s$. For these parameters, there exists a \emph{flat chimera} (limit cycle) on $\Pc_{1}\cup\Pc_2$ that consists of points with nontrivial isotropy as the plane is invariant under a reflection:
Consider the stable limit cycle shown in Fig.~\ref{fig:P3_0}(k--n) within~$\Pc_\sigma$. A numerical calculation shows that it is also stable transverse to~$\Pc_\sigma$. This limit cycle loses (transverse) stability in a pitchfork bifurcation of limit cycles which leads to two stable limit cycles with trivial isotropy that are mapped to each other through the reflectional symmetry. (Note that other stable limit cycles on~$\Pc_\sigma$, such as those shown in Fig.~\ref{fig:P3_0}(g), which are unstable transversely and therefore lead to saddle limit cycles in a pitchfork bifurcation.) Since the bifurcations happen simultaneously on~$\Pc_1$ and~$\Pc_2$, there is a total of four such solutions.
The projections of four chimera states into the $(\psi_1, \psi_3)$-plane are shown in Fig.~\ref{fig:B-M_Chaos-Proj}(a). Note that each of the resulting limit cycles has a setwise reflectional symmetry. This symmetry is broken in a supercritical pitchfork bifurcation of limit cycles at $\alpha_\s \approx 1.616$, which leads to a total of eight nonsymmetric weak chimera limit cycle solutions; cf.~Fig.~\ref{fig:B-M_Chaos-Proj}(a,b)

Each stable limit cycle loses stability in a period doubling bifurcation; see Fig.~\ref{fig:B-M_Chaos-Proj}(b--c). The resulting limit cycle bounds a M\"{o}bius strip that wraps around the torus in the~$\psi_2$ direction; the original limit cycle is of saddle type after the bifurcation and lies at the center of the strip.
Note that the width of the M\"{o}bius strip varies along variable~$\psi_2$ and for different parameter values. This indicates the heterogeneity of the attraction strength along the periodic solution that will have a further impact on the formation of chaos.
The first period doubling bifurcation is followed by a chain of period-doubling bifurcations: The $n$th period doubling bifurcation leads to a $2^n$-limit cycle as shown in Fig.~\ref{fig:B-M_Chaos-Proj}(c--f). At each period doubling bifurcation the geometry of the M\"obius strips becomes more elaborate, allowing trajectories to follow different directions as the trajectory wraps around the torus in $\psi_2$ direction.

\begin{figure}
\begin{center}
 \includegraphics[width=\linewidth]{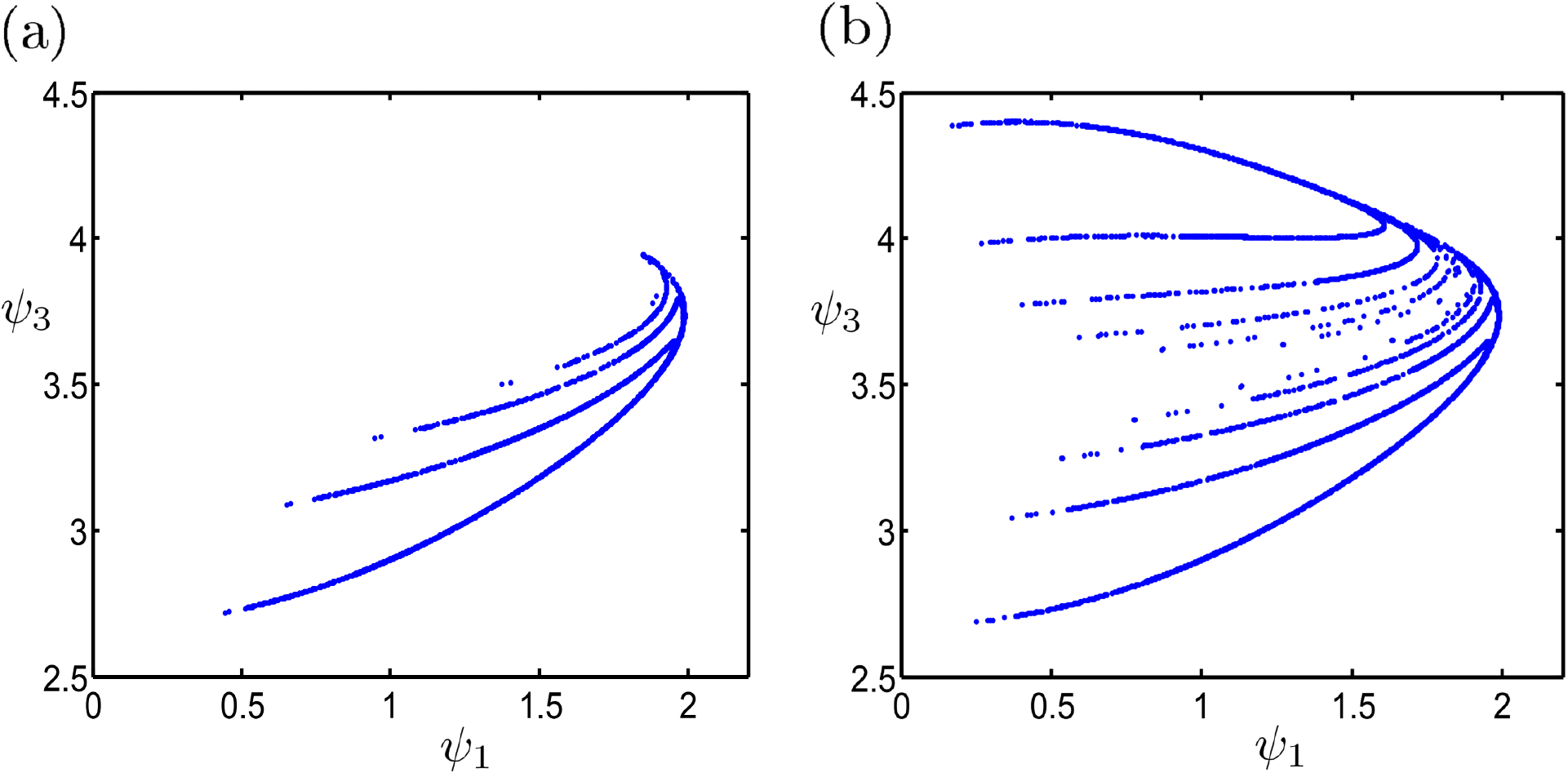}
 \end{center}
 \caption{\label{fig:B-M_3D-Chaos}
 Poincar\'e sections show the finer structure of the chaotic attractors that arise in~\eqref{eq:goveqn}. 
 Panel~(a) shows the attractor in Fig.~\ref{fig:B-M_Chaos-Proj}(g).
 Panel~(b) shows the attractor in Fig.~\ref{fig:B-M_Chaos-Proj}(h).
}
\end{figure}

A chaotic attractor with nontrivial winding number {(i.e.,} a chaotic weak chimera{)} emerges as a result of a period-doubling cascade as shown in Fig.~\ref{fig:B-M_Chaos-Proj}(g). We estimate the fractal dimension of the chaotic attractor to be slightly larger than two.
Roughly speaking, as trajectories wind around~$\psi_2$ on the attractor, they can either take a direct path, closer to the original limit cycle, or make an `excursion' in the~$\psi_1$ or~$\psi_3$ direction.
The Poincar\'e section shown in Fig.~\ref{fig:B-M_3D-Chaos}(a) shows the finer structure of the attractor.
The attractor undergoes a symmetry increasing bifurcation as two symmetrically related attractors merge (Fig.~\ref{fig:B-M_Chaos-Proj}(g--h)); a similar effect was recently observed in coupled Stuart-Landau oscillators with symmetry~\cite{Haugland2021}. The enlargement of the chaotic attractor can also be seen in the Poincar\'e section in Fig.~\ref{fig:B-M_3D-Chaos}(b); note that the section does not necessarily respect the attractor symmetry.

As the parameter~$\alpha_\s$ is increased beyond $\alpha_\s=1.64166$,
there is multistability between the equilibrium~$\SSzero$ and 
the four stable symmetric chaotic weak chimeras. 
Moreover, there are two saddle equilibria~$S_\ell$ on each of the invariant planes~$\Pc_1, \Pc_2$, as shown in Fig.~\ref{fig:P3_0}(h) or Fig.~\ref{fig:P3_0}(m). The stable manifold 
of each saddle is two-dimensional, intersect the appropriate invariant plane transversely, and  separates the basin of attraction of~$\SSzero$ and the chaotic attractors. Finally, the chaotic attractor is destroyed through a homoclinic
tangency~\cite{Lukyanov1978,Gonchenko_1997,Bonatti-Diaz-Viana,Delshams_2015} and trajectories eventually converge to~$\SSzero$.
The transients in the vicinity of the former chaotic attractor can be very long before the stable equilibrium is approached.

\subsection{Minichimerapedia for networks of two populations of two phase oscillators}

To summarize, we {have} given an overview of the solutions of~\eqref{eq:goveqn} that correspond to weak chimeras for two coupled populations of phase oscillators~\ref{eq:goveqfull}. All of these solutions share the property that the phase difference between the populations ($\psi_2$ in the reduced dynamics) is unbounded as time evolves. As they wrap around the torus, the solutions are nonhomological to zero and must arise in global bifurcations as described above.

\begin{enumerate}
\item Limit cycle solutions on the invariant planes $\Pc_1=\sset{\psi_1=0}$ or $\Pc_2=\sset{\psi_3=0}$ (Fig.~\ref{fig:P3_0}(k--n)), referred to as {flat chimeras}. The situation corresponds to phase synchronization of one of the populations with local order parameter $\abs{Z_1}=1$ (or $\abs{Z_2}=1$) and $\abs{Z_2}=\abs{Z_2(t)}\in(0,1)$ (or $\abs{Z_1}=\abs{Z_1}(t)\in(0,1)$).
\item A one-parameter family of periodic orbits on the invariant plane for $\alpha_\s=\pm\frac{\pi}{2}$, $\alpha_\n=0$, $\alpha_\n=\pm\frac{\pi}{2}$, $\alpha_\n=\pi$ (Fig.~\ref{fig:B-M_Cons-Dis}(a)).\label{itm:Same}
\item The limit cycles with setwise symmetry (Fig.~\ref{fig:B-M_Chaos-Proj}(a)).
\item The limit cycles without symmetry (Fig.~\ref{fig:B-M_Chaos-Proj}(b--f)). 
\item A two-parameter family of neutrally stable periodic orbits (parameters as in item~\ref{itm:Same}); cf.~Fig.~\ref{fig:Cube_Psi-var}(c), Fig.~\ref{fig:Dyn_On_Surf}(a,d).
\item The eight nonsymmetric chaotic attractors (Fig.~\ref{fig:B-M_Chaos-Proj}(g), Fig.~\ref{fig:B-M_3D-Chaos}(a),(c)).
\item The four symmetric chaotic attractor that appear as the result of the symmetry increasing   bifurcation (Fig.~\ref{fig:B-M_Chaos-Proj}(h), Fig.~\ref{fig:B-M_3D-Chaos}(b,d)).
\end{enumerate}
Note that trajectories close to homoclinic/heteroclinic orbits that are nonhomological to zero (see Fig.~\ref{fig:P3_0}(f)) can also show (transient) frequency-unlocked dynamics between the populations

\section{Coexistence of conservative and dissipative dynamics} 

\label{sec:ConservativeDissipative}

Finally, we remark that the system also has an interesting and unusual dynamics {when} $\alpha_\s=\pm\frac{\pi}{2}$, $\alpha_\n\in\sset{0,\pi}$---this corresponds to the function of the coupling function~$g_\s$ that determines the coupling with populations being even and the interaction function~$g_\n$ between populations being odd. In this case, the system has the first integral given in Proposition~\ref{prop:HmO}; for simplicity, we use the same notation here.
\begin{prop}
For $\alpha_\s=\frac{\pi}{2}$, $\alpha_\n=0$ and arbitrary~$A$ the system has the first integral 
$H^{(-1,0)}$
as defined in~\eqref{eq:HmO}.
\end{prop}
Thus, the dynamics of the system evolve on the level sets $H^{(-1,0)}(\psi_1,\psi_3)=C$ for fixed $C\in [-1,1]$; see Fig.~\ref{fig:First_Int_A}(b).
In the cases $C=\pm 1$ we have a planar system on the invariant planes $\Pc_1=\sset{\psi_3=0}$ and $\Pc_2=\sset{\psi_1=0}$.
As $\abs{C}$ is decreased, the level sets deform continuously and limit to dynamics on the half-planes $\psi_3=\psi_1$ for $\psi_1\in[0,\pi]$ and $\psi_3=2\pi-\psi_1$ for $\psi_1\in[\pi,2\pi]$ (or their symmetric counterparts) for $C=0$.

This system has coexistence of conservative (Hamiltonian-like) and dissipative dynamics in phase space for $\alpha_\s=\frac{\pi}{2}$, $\alpha_\n=0$, and $A\in[A^*,1/A^*]$ with $A^*\approx 0.38146$.
Conservative-dissipative dynamics are often related to the presence of time-reversing symmetries; such dynamics has been described, for example, in a three-dimensional laser system~\cite{Politi1986}, globally coupled superconducting Josephson junction arrays~\cite{Tsang1991}, chains of locally coupled phase oscillators~\cite{Topaj-Pikovsky}, and circulant networks of phase oscillators with skew-symmetric coupling~\cite{BMWY_SIAM_2018}. 
Here, the system has the conservative region filled with a two-parameter family of neutral periodic orbits (weak chimeras) and this region is bounded with the surface of one-parameter family of heteroclinic cycles forming a ``heteroclinic tube''.
Outside the Hamiltonian-like region, the dynamics is dissipative with attractor~$\SSzero$ and repeller~$\SSpi$ when $A<1$ or attractor~$\SSpi$ and repeller~$\SSzero$ when $A>1$.
The boundary surface of the 3-dimensional conservative region has a structure similar to that shown in Fig.~\ref{fig:Chimera-Snake}(a).
The system has two one-parameter families of saddle points (which are curved lines compared to the case $\alpha_\s=\alpha_\n=\pm\frac{\pi}{2}$). There are heteroclinic (homoclinic in~$\Tor^3$) cycles that consist of these saddles and their stable and unstable invariant manifolds on the same level surface (each saddle is neutral in transversal towards the level surface). 
Note that while for fixed $\alpha_\s=\frac{\pi}{2}$ the dynamics for $\alpha_\n=\frac{\pi}{2}$ (see above) and $\alpha_\n=0$ both take place on neutrally stable cylindrical surfaces, these surfaces have essentially different shapes (compare Fig.~\ref{fig:Cube_Psi-var}(d) and Fig.~\ref{fig:First_Int_A}(b)):
Invariant lines that consist of neutral saddles are no longer straight lines (as~$\Ic_1$ and~$\Ic_2$ for $\alpha_\n=\frac{\pi}{2}$) but they deform when changing the parameter~$A$.

\begin{figure}
\begin{center}
 \includegraphics[width=1\linewidth]{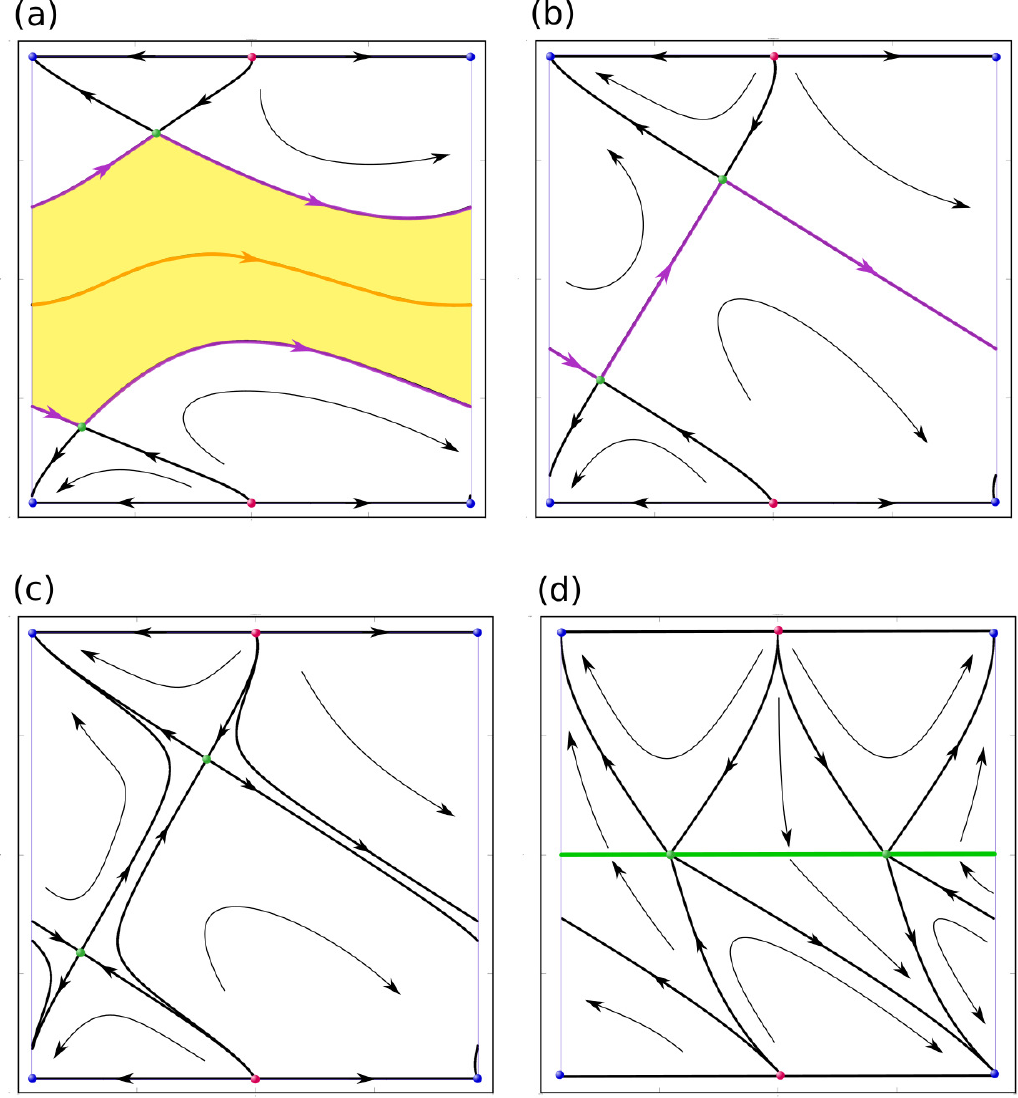}
 \end{center}
 \caption{\label{fig:B-M_Cons-Dis}
 Coexisting conservative and dissipative arise for $M=2$ populations of $N=2$ phase oscillators~\eqref{eq:goveqfull}. Here schematic phase portraits on the cylindric surfaces $H^{(-1,0)}=C$ 
are shown for variables~$(\psi_2,\phi)\in \mathbb{T}^2$, where~$\phi$ is a parametrization of the level line of the first integral. 
 Panels (a--c) show the bifurcation transition that leads to the disappearance of the conservative region (shaded in yellow) filled with periodic (chimera) trajectories. 
 Panel~(d) shows dynamics for $C=0$ when level surface consists of two half-planes; the intersection of these planes is~$\Ic_0$ (green line).
 Fixed points are colored according to type: source (red), sink (blue), saddle (green), saddle-node (two-color). Purple lines are heteroclinic orbits.
}
\end{figure}

The theory of bifurcations without parameters~\cite{Fiedler2000, Liebscher2011} can also be used to study the dynamics of the current case.
We can consider two-dimensional dynamics on surfaces $H^{(-1,0)}=C$ using~$C$ as a parameter. This allows to analyze conservative-dissipative dynamics on an individual level set and then extrapolate to the entire phase space~$\mathbb{T}^3$ as~$C$ parametrizes the level sets and attractors and repellers are located on the common boundary of all level sets $\psi_1=\psi_3=0$.
Fig.~\ref{fig:B-M_Cons-Dis} shows schematic phase portraits on surfaces $H^{(-1,0)}=C$ for different~$A$ and~$C$. The conservative region on the cylindric surface is the biggest for $C=\pm 1$ for any $A=1$ and it shrinks with decreasing~$|C|$ to the heteroclinic cycle (transition from~(a) to~(c) in Fig.~\ref{fig:B-M_Cons-Dis}). Fig.~\ref{fig:B-M_Cons-Dis}(d) corresponds to the limit case $C=0$ with degenerate (green) line~$\Ic_0$ of saddles in the entire phase space~$\mathbb{T}^3$. There are two conservative regions close to the straight lines $(0,\psi_2,\pi)$ and $(\pi,\psi_2,0)$ and also one common dissipative region in~$\Tor^3$.

For $A=1$ the three-dimensional conservative region corresponds to the whole phase space~$\Tor^3$. Finally, the conservative region is destroyed in a saddle-connection bifurcation when the connections of stable and unstable invariant manifolds of each saddle are simultaneously broken.
This happens either when one of {the} equalities $\alpha_\s=\pm\frac{\pi}{2}$, $\alpha_\n=0$ is violated or when the parameter~$A$ {leaves} the interval~$[A^*,1/A^*]$: If~$A$ reaches the bifurcation value~$A^*$ or~$1/A^*$ (Fig.~\ref{fig:P3_0}(c)), the conservative region collapses onto the one-dimensional heteroclinic cycle (between two saddles that belong to $\psi_1=0$ or $\psi_3=0$).

\section{Discussion}

Here we considered properties of phase oscillator networks that consist of $M$~populations of $N$~phase oscillators. While {we} analyzed networks for $M=N=2$ with sinusoidal coupling in details, some of the observation hold also for larger networks with more general coupling. We therefore briefly discuss a few {general} properties of~\eqref{eq:PopMN}. First, the system~\eqref{eq:PopMN} has dihedral symmetry $\Di_{MN}$ for any~$M$ and~$N$. Second, \eqref{eq:PopMN} is a gradient system for any odd coupling functions~$g_\s$, $g_\n$ for all~$M$ and~$N$. For example, if $g_\s(\phi)=g_\n(\phi)=g(\phi)$ satisfies $g(\phi)=-g(-\phi)$ then the system has {the} potential
\begin{equation*}\label{eq:Potential}
\begin{split}
V(\theta) &= - \frac{1}{2MN}\sum_{\sigma=1}^{M}\sum_{k=1}^{N}
\Big(K_\s\sum_{j=1}^N h(\theta_{\sigma,j}-\theta_{\sigma,k})\\ &\qquad\qquad\qquad\qquad\quad
+K_\n\sum_{\tau\neq \sigma}\sum_{j=1}^N h(\theta_{\tau,j}-\theta_{\sigma,k}) \Big),
\end{split}
\end{equation*}
where~$h(\phi)$ is an even function such that $h'(\phi)=g(\phi)$. The gradient system~\eqref{eq:GradDyn} is a special case. Third, in case of an even coupling function $g_\s(\phi)=g_\n(\phi)=g(\phi)$ with $g(\phi)=g(-\phi)$ the system~\eqref{eq:PopMN} is divergence-free (this generalizes results in Ref.~\onlinecite{Ashwin2016} for $K_\s=K_\n$). Other properties will be discussed elsewhere.

\begin{figure}
\begin{center}
 \includegraphics[width=\linewidth]{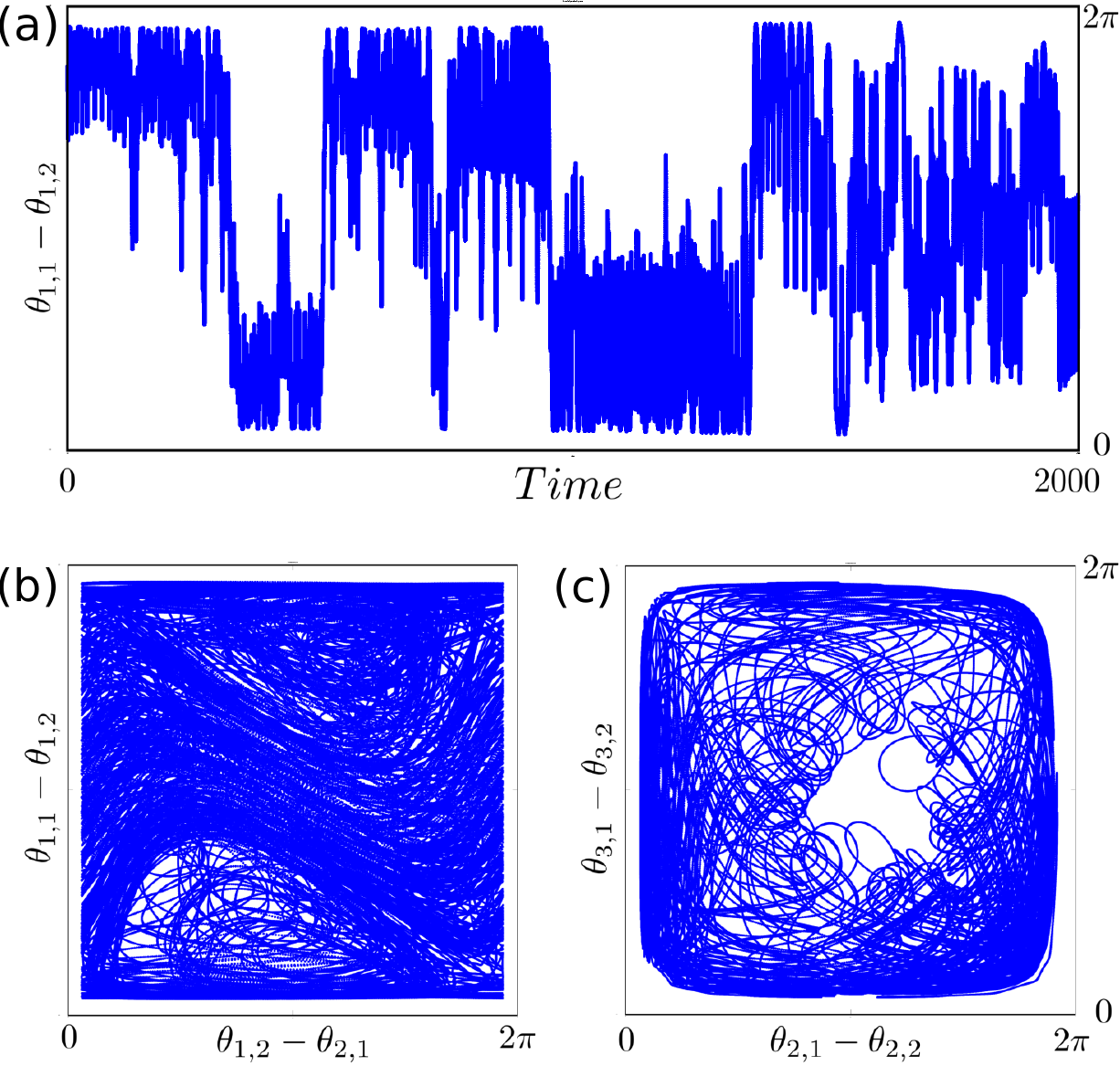}
 \end{center}
 \caption{\label{fig: B-M_3g2o_pi2_05}
Example of a neutral chaotic weak chimera in the six oscillator system for $M=3$, $N=2$ in {the} conservative case $\alpha_\s=\alpha_\n=\frac{\pi}{2}$, $A=\frac{1}{2}$. (a) Time series of phase difference between two oscillators of the first group.
(b) and (c): Projections of trajectories from $\mathbb{T}^5$ into phase planes of the phase differences. Phase differences are locked if two oscillators belong to the same group, and such differences are phase-unlocked if the oscillators belong to different groups (as $\theta_{1,2}-\theta_{2,1}$).
}
\end{figure}

Our results also shed light on the phase space structure in higher dimensions, for example, $M=3$ populations of $N=2$ oscillators.
In this case, the system has a continuous set of neutral chimera solutions for coupling $g_\s(x)=g_\n(x)=\pm\cos(x)$ (or $\alpha_\s=\alpha_\n=\pm\frac{\pi}{2}$). These solutions can be periodic, quasi-periodic or chaotic (similar to ABC flows~\cite{Arnold(ABC),Dombre1986}, chaos that fills an entire torus). Fig.~\ref{fig: B-M_3g2o_pi2_05} shows an example of such chaotic dynamics for $M=3$ populations of $N=2$ oscillators.
This suggests that the situation with $(N-2)$-parametric neutral chimeras also occurs in the case of arbitrary~$M$, $N$ for $\alpha_\s=\alpha_\n=\pm\frac{\pi}{2}$.

The system in~\eqref{eq:PopMN} with $M=2$ populations and $N>3$ oscillators has been subject of a number of studies~\cite{Panaggio2015} with small and large finite oscillator systems~\cite{Montbrio2004,Panaggio2016,Bick2018}, including the continuum limit~\cite{Abrams2008,Martens2016,Martens2016b,Bick2018} of infinitely many oscillators,
some based on low-dimensional descriptions obtained via the Watanabe--Strogatz or Ott--Antonsen reductions~\cite{Watanabe1994,Ott2008,Bick2018c}.  
For future work, it would be interesting to investigate how our findings extend to the dynamic behavior of such larger system sizes, particularly regarding invariant subspaces and how they are affected by the symmetry breaking mechanism leading to weak chimera states.

Previous studies found that bifurcation curves leading to chimera states are organized around points in parameter space where $|\alpha|=\frac{\pi}{2}$~\cite{Ashwin2014a,Martens2016,Bick2017}; here, we explained the symmetry breaking bifurcations generating weak chimeras near these points.
It is interesting to note that experiments using mechanical oscillators~\cite{MartensThutupalli2013,Panaggio2015} indicate that chimera states emergence in a scenario near resonance---this conjecture is confirmed as one can rigorously show~\cite{MartensThutupalli2022} that resonance in such systems is related to phase lags being $\alpha_{\s},\alpha_\n=\pm\frac{\pi}{2}$.

\section*{Acknowledgements}
O.B. acknowledges the support of the National Research Foundation of Ukraine (Project number 2020.02/0089).

\appendix


\bibliographystyle{unsrt}

\end{document}